\documentclass[rmp,onecolumn,notitlepage]{revtex4-1}
\usepackage[a4paper,top=2.00cm,bottom=2.00cm,left=2.00cm,right=2.00cm]{geometry}
\usepackage{amscd}
\usepackage{amsfonts}
\usepackage{amstext}
\usepackage{amsmath}
\usepackage{amssymb}
\usepackage{mathptmx}
\usepackage{bm}
\usepackage{esint}
\usepackage{enumerate}
\usepackage{float}
\usepackage{lastpage}
\usepackage[utf8]{inputenc}
\fontfamily{cmr}\selectfont
\usepackage{graphicx}
\DeclareGraphicsExtensions{.eps, .pdf, .jpg, .png}
\makeatletter
\def\lastpage@putlabel{}
\makeatother
\usepackage[colorlinks=true,linkcolor=blue,citecolor=blue]{hyperref}
\usepackage{color}
\graphicspath{{./}}
\renewcommand{\onlinecite}[1]{\hspace{-1 ex} \nocite{#1}\citenum{#1}}
\usepackage{footmisc}
\usepackage{natbib}
\setcitestyle{super}
\usepackage[force]{feynmp-auto}
\usepackage{bookmath}
\usepackage{mathfrak}
\usepackage{mymatrices}
\usepackage{mygroups}
\setcitestyle{numbers,super}
\date{\today}
\begin{document}
\title{Theory of Disordered Superconductors \\ 
       \qquad{\sl with Applications to Nonlinear Current Response}
}
\author{J. A. Sauls} 
\email{sauls@northwestern.edu}
\address{Center for Applied Physics \& Superconducting Technologies 
\\
Department of Physics \& Astronomy,
\\
Northwestern University,
\\
Evanston, IL 60208 USA}
\begin{abstract}
I present a review of the theory and basic equations for charge transport in superconducting alloys starting from the Keldysh formulation of the quasiclassical transport equations developed by Eilenberger, Larkin and Ovchinnikov and Eliashberg. This formulation is the natural extension of Landau's theory of normal Fermi liquids to the superconducting state of strongly correlated metals. 
For dirty metals the transport equations reduce to equations for charge diffusion, with the current response given by the Drude conductivity at low temperatures.
The extension of the diffusion equation for the charge and current response of a strongly disordered normal metal to the superconducting state yields Usadel's equations for the non-equilibrium quasiclassical Keldysh propagator.
The conditions for the applicability of the Usadel equations are discussed, the pair-breaking effect of disorder on the current response, including the nonlinear current response to an EM field in the dirty limit, $\tau \ll \hbar/\Delta$, are reported. The same nonlinearity is shown to lead to source currents for photon generation and nonlinear Kerr rotation driven by the nonlinear response to excitation of the superconductor by a multi-mode EM field. The potential relevance of the nonlinear source currents to SRF cavities as detectors of axion-like dark matter candidates is briefly discussed.
\end{abstract}
\maketitle
\section{Introduction}

\begin{figure}[t]
\begin{minipage}{\columnwidth}
\input{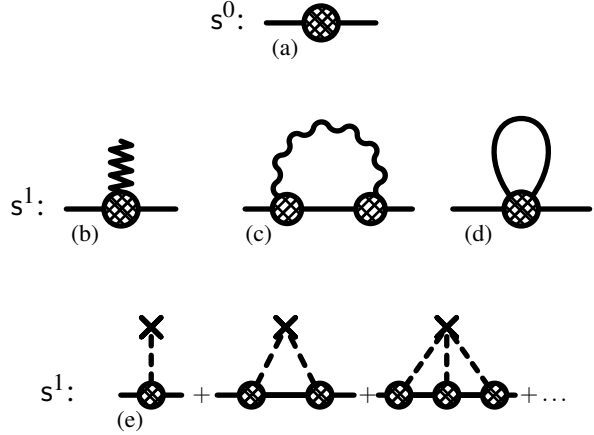}
\caption{Leading order electronic self-energy diagrams of the Fermi-liquid theory of superconductivity. These terms are first order in the expansion parameter \sml. The diagrams describe:
      (a) the zeroth-order self energy defines the Fermi surface and Fermi velocity,
      (b) the coupling of an external field to low-energy quasiparticles,
      (c) Migdal's leading-order quasiparticle-phonon self energy,
      (d) the mean-field interaction energy of quasiparticles and Cooper pairs, and
      (e) the leading-order quasiparticle-impurity scattering self energy diagrams.
}
\label{fig-Sigma-small}
\end{minipage}
\end{figure}

Strongly interacting Fermions can form a ``Fermi-liquid state'' in which the physical properties at low temperatures are dominated by low-lying excitations (quasiparticles) which are composite objects but have basic features (e.g. charge, spin, fermion number) in common with non-interacting electrons.\cite{lan56} 
At the heart of Landau's theory is the distribution function $n(\vp,\vr;\varepsilon,t)$,\cite{lan57}$^{,}$ describing the dynamics of an ensemble of quasiparticles in phase space ($\vp,\vr$) governed by the Boltzmann-Landau transport equation.
Derivations of this transport equation from first principles use many-body Green's function techniques, and lead to explicit expressions for the various terms of the transport equation in terms of self-energies.\cite{eli62,pra64} The self-energies describe the effects of electron-electron, electron-phonon and electron-impurity scattering. The set of Feynman diagrams for the relevant self-energies are shown in Fig. \ref{fig-Sigma-small}. Note that the hatched circles are block vertices representing renormalized quasiparticle-quasiparticle, quasiparticle-phonon, and quasiparticle-impurity interactions. These interactions, together with the Fermi-surface properties (i.e. Fermi momentum, $\vp$, Fermi velocity, $\vv_{\vp}$, topology of the Fermi surface), can in principle be calculated from the full many-body theory, but often are treated as parameters of the Fermi-liquid theory, and are obtained by comparison of theory with experiment.
Landau's theory predicts a number of universal results for temperature and magnetic field dependences of thermodynamic and transport properties at low temperature. The universal laws can provide signatures for Fermi-liquid behaviour. A detailed discussion of these normal-state properties can be found in various review articles and textbooks, c.f. Ref.~\onlinecite{baym91}.

The classical phase-space structure of Landau's Fermi-liquid theory for normal metals is absent in the microscopic theory of superconductivity by Bardeen, Cooper and Schrieffer (BCS),\cite{bar57} and the quantum field theory formulations by Bogoliubov\cite{bog58a} and Gorkov.\cite{gor59}
It was more than ten years after the BCS breakthrough before Landau's theory of Fermi-liquids and BCS theory were cast into a common theoretical framework. 
The \emph{quasiclassical} theory of superconductivity was formulated in a series of publications starting with Eilenberger's reduction of Gorkov's equations to transport-like equations for the normal and anomalous Greens functions for equilibrium states of type II superconductors.\cite{eil68} Larkin and Ovchinnikov independently derived the quasiclassical transport equations.~\cite{lar69} Eliashberg,~\cite{eli72} and Larkin and Ovchinnikov~\cite{lar75} generalized the quasiclassical theory to nonequilibrium states of superconductors with strong electron-phonon interactions.
This theory allows one to calculate essentially all superconducting phenomena of interest, from transition temperatures, excitation spectra, Josephson effects, vortex structures, to the response of superconductors to electromagnetic fields.\cite{rai94} 
In quasiclassical theory the dynamics of quasiparticles is described partly by classical statistical mechanics, and partly by quantum mechanics. The classical degrees of freedom are the motion of quasiparticles in $\vp$-$\vr$ phase space; i.e. quasiparticle wavepackets move along classical trajectories. Quantum coherence between particle and hole states is the key quantum mechanical degree of freedom encoded in the BCS theory of superconductivity, and is the origin of the non-classical phenomena associated with the superconducting state, e.g. persistent currents, perfect diamagnetism, the Josephson effect and branch-conversion (Andreev) scattering.
Particle-hole coherence is described in the quasiclassical theory by grouping particle excitations, occupied one-electron states with energy above the Fermi energy ($\varepsilon >0$), and hole excitations, empty one-electron states with $\varepsilon <0$, into an iso-spin doublet. The particle-hole doublets (Nambu spinors) span a two-dimensional space of quasiparticle excitations (Nambu space). The quantum statistics of the internal state of quasiparticle excitations is described by a $2\times 2$ density matrix for the particle-hole degree of freedom.\footnote{I consider spin-singlet superconductors and neglect spin-orbit interactions and paramagnetism, in which case the $4$-dimensional density matrices describing the combined spin and particle-hole degrees of freedom reduce to $2\times 2$ matrices in particle-hole space.\label{foot-singlet}}

Here I focus on the theory of superconductivity in metals with impurity disorder in the ``dirty limit'' defined by electron-impurity scattering mean free paths short compared to the superconducting coherence length, $\ell \ll \xi$. As an introduction, Sec.~\ref{sec-Normal_State-Charge_Diffusion}, I discuss the reduction of the nonequilibrium transport equations for disordered normal metals to a transport equation for the Keldysh propagator, and the corresponding charge response functions. 
In Sec.~\ref{sec-Usadel_Equations} I develop the the non-equilibrium Keldysh-Eilenberger equations as an expansion in the parameter, $\delta\equiv\ell/\xi$, and obtain non-equilibrium quasiclassical equations as a generalization of Usadel's equations for equilibrium states of inhomogeneous superconductors.~\cite{usa70} For a complementary development of the non-equilibrium Usadel equations based on a functional integral formulation see Ref.~\onlinecite{kamenev11}.
The sensitivity of conventional ``s-wave'' superconductors to disorder is discussed in the context of the Usadel equations, highlighting the insensitivity of the transition temperature and excitation gap to disorder (Sec.~\ref{sec-Anderson_Theorem}) in contrast to the strong suppression of the supercurrent to impurity disorder (Sec.~\ref{sec-Current_Response}).
In Sec.~\ref{sec-Nonlinear_Currents} I derive the nonlinear current response to the gauge-invariant condensate momentum, $\vp_s=\nicefrac{1}{2}(\gradr\vartheta-\nicefrac{2e}{c}\vA)$, and in Sec.~\ref{sec-Nonlinear_Meissner_Effect} obtain the field-dependence of the London penetration depth from the nonlinear screening currents (\emph{nonlinear Meissner effect}) at a vacuum-superconducting interface with a external field parallel to the interface.
In Secs.~\ref{sec-Microwave_Photon_Generation}-\ref{sec-Nonlinear_Kerr_Effect} the nonlinear Meissner response is extended to microwave frequencies. The nonlinear response leads to a number of novel effects including photon generation at the third harmonic, and photons at intermodulation frequencies for multi-mode excitation of the superconductor, as well as nonlinear Kerr rotation. In Sec.~\ref{sec-Nonlinear_SRF_Signal} I discuss the potential relevance of the nonlinear current response to SRF cavities as possible detectors of axion-like dark matter.  
I begin with a review of the quasiclassical transport theory based on the leading order self energies for disordered superconductors.

\section{Theory}\label{sec-Quasiclassical_Theory}

The central objects of the theory are the propagators, $\whmfGrak(\vp,\vr;\varepsilon,t)$, where the superscript on the propagator identifies its microscopic significance; $\whmfGk$ denotes the Keldysh propagator, which is a $4\times 4$ matrix generalization of the classical Boltzmann-Landau distribution function to the superconducting state.\footnote{In the ``energy representation'' the traditional variables of the distribution function, $\{\vp$, $\vr$, $t\}$, are replaced by the equivalent set $\{\vp_f, \varepsilon,\vr,t\}$. The three-dimensional momentum variable $\vp$ is replaced by the two-dimensional Fermi momentum, $\vp_f$, which is defined by the direction normal to the Fermi surface for the momentum $\vp$ nearest to $\vp_f$, and the excitation energy, $\varepsilon=E(\vp,\vr;t)-E_f$. For a review of transport theory, including the energy representation, see Refs. (\onlinecite{ram86,rai94}). I use the energy representation throughout this paper, in which case I drop the subscript on the momentum variable, so hereafter $\vp$ is understood to be a value of the momentum on the Fermi surface.} 
The matrix structure encodes the spin and particle-hole isospin degrees of freedom, The matrix elements of $\whmfGk$ in particle-hole space are,
\begin{equation}\label{eq-GK} 
\whmfGk=
\left(
\begin{array}{cc}
\hmfgk & \hmffk \\
\hmfufk & \hmfugk
\end{array}\right)
\,,
\end{equation}
where the off-diagonal element, $\hmffk(\vp,\vr;\varepsilon,t)$, is the \emph{anomalous} (Gorkov) propagator, which encodes the dynamics of Cooper pairs and is the source of particle-hole coherence. For the case of strong disorder only conventional isotropic spin-singlet pairing survives the random potential. In this case the spin structure of the anomalous propragators reduces to $\hmffrak=i\sigma_y\,\mffrak$, where $\sigma_y$ is the anti-symmetric Pauli spin matrix.~\footref{foot-singlet}
The conjugate anomalous propagator, $\hmfufk=i\sigma_y\,\mfufk$, is related to $\hmffk$ by particle-hole conjugation symmetry, $\mfufk(\vp,\vr;\varepsilon,t) = -\mffk(-\vp,\vr;-\varepsilon,t)^*$. All of the relevant propagators and the symmetry relations connecting them are summarized in Appendix~\ref{appendix-propagators}

The diagonal elements, $\hmfgrak=\mfgrak\,\hone$ and $\hmfugrak=\mfugrak\,\hone$, determine the distribution functions, $n_p$ and $n_h$, for particle and hole excitations, respectively\footnote{$n_{p(h)}(\vp,\vr;\varepsilon,t)\,d\varepsilon\,d^2\vp\,d^3R$ is the number of particle excitations (hole excitations) with excitation energy $\varepsilon$, momentum $\vp$ on the Fermi surface, and position $\vr$ in the phase space element $d\varepsilon\,d^2\vp\,d^3R$.}
\ber
\label{eq-particle_distribution}
\frac{1}{4\pi i}
\left(\mfgk-(\mfgr-\mfga)\right)
&=&
n_p\,N(\vp,\vr;\varepsilon,t)
 \,, 
\\
\frac{1}{4\pi i}
\left(\mfugk - (\mfugr -\mfuga)\right)
&=&
n_h\,N(\vp,\vr;\varepsilon,t)
\,,
\label{eq-hole_distribution}
\eer
where $\mfgra$ and $\mfugra$ are the corresponding diagonal components of the retarded (R) and advanced (A) quasiclassical propagators, $\whmfGra$, which determine the local space-time-dependent quasiparticle spectral function,
\be\label{eq-LDOS}
N(\vp,\vr;\varepsilon,t)=N_f\,\left[-\frac{1}{2\pi}\Im\,\mfgr(\vp,\vr;\varepsilon,t)\right]
\,,
\ee
where $N_f=m^* p_f/2\pi^2\hbar^3$ is the density of states of the normal metal at the Fermi energy, $p_f$ ($v_f$) is the magnitude of the Fermi momentum (velocity), and $m^*=p_f/v_f$ is the quasiparticle effective mass.
For equilibrium states at temperature $T$ the Keldysh propagator is given by $\mfgk=\tanh(\beta\varepsilon/2)\,\left[\mfgr-\mfga\right]$, in which case the particle and hole distribution functions reduce to the corresponding Fermi distributions, $n_p^{\text{eq}} = f(\varepsilon) = 1/(e^{\beta\varepsilon} + 1)$, $n_h^{\text{eq}} = f(-\varepsilon) = 1 - f(\varepsilon)$, and $\beta=1/\kb T$.

Measureable properties such as the charge density, $n(\vr,t)$, and charge-current density, $\vj(\vr,t)$, are obtained from the diagonal components of the quasiclassical Keldysh propagator.
If I neglect the Landau mean-field self energy term (diagonal contribution from diagram \ref{fig-Sigma-small}(d)), the charge density is given by
\be\label{eq-charge_density} 
\hspace*{-3mm}
n(\vr,t)
\ns=\ns 
n^{\text{l.e.}}(\vr,t)
\ns+ \ns
2eN_f\ns\int\ns d^2\vp\ns\int\ns\frac{d\varepsilon}{4\pi i}\mfgk(\vp,\vr;\varepsilon,t)
\,,
\hspace*{3mm}
\ee
where $n^{\text{l.e.}}=n_0+ 2e^2N_f\Phi(\vr,t)$ is the local equilibrium charge density of electrons in the presence of an electro-chemical potential $\Phi$. The integral over the Keldysh propagator is the charge density from deviations from local equilibirum. The corresponding current density is given by
\be\label{eq-current_density}
\vj(\vr,t) = 2eN_f\int d^2\vp \int\frac{d\varepsilon}{4\pi i}\,\vv_{\vp}\,\mfgk(\vp,\vr;\varepsilon,t)
\,,
\ee
where $\vv_{\vp}$ is the Fermi velocity. 
Equations \ref{eq-particle_distribution} and \ref{eq-hole_distribution} can be used to separate the charge density and charge-current density into contributions from the particle- and hole distribution functions, $n_p$ and $n_h$, and the spectral functions, $\mfgr-\mfga$ and $\mfugr-\mfuga$. The charge and current densities include contributions from particle and hole excitations, represented by $n_p$ and $n_h$, and contributions from the superconducting condensate, represented by the non-equilibrium spectral functions. Note that for spatially varying superconducting states the local equilibrium supercurrents are encoded in the Keldysh propagator. 

The central equations of the nonequilibrium theory of superconductivity are transport-like equations for $\whmfGrak$, which generalize the Landau-Boltzmann equation to the superconducting state.\cite{eil68,lar75} They are a set of coupled integral-differential equations of first order in the spatial derivatives and, in general, of infinite order in time derivatives. The equation for the Keldysh propagator is,
\begin{eqnarray}
 \whHamr\circ\whmfGk(\vp,\vr;\varepsilon,t)
-\whmfGk\circ\whHama(\vp,\vr;\varepsilon,t)
+\whmfGr\circ\whSigk(\vp,\vr;\varepsilon,t) 
-\whSigk\circ\whmfGa(\vp,\vr;\varepsilon,t)
+i\hbar\vv_{\vp}\cdot\gradr\whmfGk(\vp,\vr;\varepsilon,t) 
= 0
\,,
\label{eq-QC_K}
\end{eqnarray}
where the operators 
\begin{equation}\label{eq-QC-operator}
\whHamra(\vp,\varepsilon;\vr,t)
\equiv
\varepsilon\tz-\whv(\vp;\vr,t)-\whSigra(\vp,\varepsilon;\vr,t)
\,,
\end{equation}
are defined by the excitation energy, $\varepsilon$, the coupling to external fields, $\whv$, and the retarded and advanced self-energies, $\whSigra$.
The directional derivative corresponds to propagation along classical trajectories defined by the Fermi velocity, $\vv_{\vp}$, and arises from the quasiclassical approximation to the inhomogeneous Gorkov equations which takes advantage of the separation in scale for spatial variations of the superconducting order parameter, set by the Cooper pair correlation length, $\xi_0\equiv\hbar v_f/2\pi\kb T_c$, and the atomic scale set by the Fermi wavelength, $\lambda_f=\hbar/p_f$, i.e. the derivative term is of order $\lambda_f/\xi_0$ with corrections of order $(\lambda_f/\xi_0)^2$ or smaller. For a more detailed discussion of the quasiclassical approximation c.f. Ref.~\onlinecite{sau18}.  

The $\circ$-product represents $2\times 2$-matrix multiplication in Nambu space combined with a convolution product in the energy-time variables defined by 
\be
\whmfa\circ\whmfb(\varepsilon,t)
=\exp
\left[
\nicefrac{i}{2}
(\partial_{\varepsilon}^{\mfa}\partial_{t}^{\mfb}-\partial_{t}^{\mfa}\partial_{\varepsilon}^{\mfb})
\right]
\whmfa(\varepsilon,t)\whmfb(\varepsilon,t)
\,,
\ee
where the superscripts $\mfa$ ($\mfb$) on the partial derivatives indicate derivatives with respect to the arguments of $\whmfa$ ($\whmfb$).
The solution of the Eq. \ref{eq-QC_K} for $\whmfGk$ requires the external potentials $\whv(\vp,\vr;t)$,
the advanced, retarded and Keldysh self-energies, $\whSigrak(\vp,\vr;\varepsilon,t)$, and the advanced and retarded quasiclassical propagators, $\whmfGra$. The latter are solutions of the retarded and advanced Eilenberger equations, 
\begin{equation}\label{eq-QC_RA}
\left[\whHamra\,,\,\whmfGra\right]_{\circ}
 +i\hbar\vv_{\vp}\cdot\gradr\whmfGra(\vp,\varepsilon;\vr,t) = 0
\,.
\end{equation}
In addition, the physical solutions of Eqs. \ref{eq-QC_K}-\ref{eq-QC_RA} satisfy the normalization conditions,
\be\label{eq-QC_normRA}
\whmfGra\circ\whmfGra = -\pi^2\,\tone
\,,
\ee
\be\label{eq-QC_normK}
\whmfGr\circ\whmfGk + \whmfGk\circ\whmfGa = 0 
\,.
\ee
The normalization condition was dervived by Eilenberger for the equilibrium propagators,\cite{eil68} and is the contraint that enforces the normalization provided by the source term of Gorkov's equation. The extension of the normalization conditions to nonequilibrium retarded, advanced and Keldysh propagators was done by Larkin and Ovchinnikov.\cite{lar75} In particular, Eq.~\eqref{eq-QC_normK} provides the starting point for transforming the Keldysh propagator into non-equilibrium distribution functions for Bogoliubov quasiparticles for long-wavelength, low-frequency disturbances from equilibrium.~\cite{ser83}

The quasiclassical transport equations, Eqs. \ref{eq-QC_K}-\ref{eq-QC_RA}, require as inputs the the self-energies, $\whSigrak$, which are defined by the re-summed perturbation expansion in terms of renormalized electron-electron, electron-phonon and electron-impurity interactions.\cite{rai86,rai94} The leading order self-energies are represented in terms of Feynman diagrams in Fig. \ref{fig-Sigma-small}. Each self energy is defined in terms of renormaized interactions, shown as block vertices which couple to the low-energy quasiclassical propagators. The self energies are functionals of the quasiclassical propgators, and are computed self consistently with the propagators obtained as solutions of Eqs. \ref{eq-QC_K}-\ref{eq-QC_normK}.
The zeroth-order diagram, Fig.\ref{fig-Sigma-small}(a), is independent of the quasiclassical propagators and represents the bandstructure self-energy; it is parametrized by the Fermi momentum, $\vp$, Fermi velocity $\vv_{\vp}$ and quasiparticle spectral weight, $a(\vp) \equiv [1-\partial\Re\Sigma^{(a)}(\vp,\varepsilon)/\partial\varepsilon|_{\varepsilon=0}]^{-1}$. The latter is absorbed into the definition of the renormalized vertices.~\cite{rai94}
The first-order self-energies shown in Fig. \ref{fig-Sigma-small} represent 
(b) the coupling of an external field to low-energy quasiparticles, 
(c) the leading-order electron-phonon self energy,
(d) the mean-field electron-electron interaction energy of quasiparticles and Cooper pairs, and 
(e) the electron-impurity scattering processes.

Here I consider the combined electron-electron and electron-phonon self energies, Figs.~\ref{fig-Sigma-small}(c)-(d), in the ``weak-coupling limit'', $\kb T_c \ll \varepsilon_c \ll E_f$, where $\varepsilon_c$ is the bandwidth of attraction for Cooper pair formation. For conventional electron-phonon mediated superconductivity, the bandwidth is determined by the maximum phonon energy, or ``Debye energy'', $\varepsilon_c\gtrsim\hbar\Omega_{\mbox{\small D}}$. The corresponding self energies are independent of energy in the low-energy bandwidth, in which case $\whSigr=\whSiga\equiv\whSig_{\mbox{\tiny mf}}(\vp,\vr;t) = \mfsig_{\text{mf}}(\vp,\vr;t)\whtauz + \whDelta(\vp,\vr;t)$. 
The diagonal term is the Landau molecular field self energy,
\begin{equation}\label{eq-molecular_field}
\mfsig_{\mbox{\tiny mf}}(\vp,\vr;t)=
\int\,d^2\vp'\,A(\vp,\vp')\,\int\frac{d\varepsilon}{4\pi i}\,\mfgk(\vp',\vr;\varepsilon,t)
\,,
\end{equation}
where $A(\vp,\vp')$ is the forward scattering limit of the quasiparticle-quasiparticle scattering amplitude for momenta on the Fermi surface.~\cite{rai94}
The off-diagonal self-energy for s-wave, spin-singlet pairing takes the form,
\be\label{eq-Delta_Nambu}
\whDelta(\vp,\vr;t)  = 
\begin{pmatrix} 0 & i\sigma_y\,\Delta(\vr;t) \\ i\sigma_y\,\Delta^*(\vr;t)  & 0 \end{pmatrix}
\,,
\ee
with the mean-field order parameter given by 
\begin{equation}\label{eq-Delta}
\Delta(\vr;t)=g
\int d^2\vp \int\frac{d\varepsilon}{4\pi i}\,\mffk(\vp',\vr;\varepsilon)
\,,
\end{equation}
where $g=\lambda-\mu^*$ is the s-wave pairing interaction from the combined electron-phonon-electron ($\lambda$) and electron-electron ($\mu^*$) interactions in the Cooper channel in the weak-coupling limit. Note that $g>0$ corresponds to a net attractive interaction. 

The effects of impurity scattering on the properties of superconductors are determined to leading-order in $\sml = \hbar/p_f \ell$, where $\ell=v_f\tau$ is the mean-free path due to quasiparticle scattering off a random distribution of impurities, by the self-energy diagrams in Fig.~\ref{fig-Sigma-small}(e).
This series defines the quasiparticle-impurity T-matrix, $\whmftrak(\vp,\vp';\varepsilon)$, with the self-energy given by the T-matrix in the forward scattering limit,
\be\label{eq-Sigma_imp}
\whSigrak_{\text{imp}}(\vp,\vr;\varepsilon,t)=n_{\text{imp}}\,\whmftrak(\vp,\vp,\vr;\varepsilon,t)
\,,
\ee
where $n_{\text{imp}}$ is the mean impurity density, and the T-matrices are the solutions of the equations
\ber\label{eq-T-matrix_RA} 
\whmftra(\vp,\vp',\vr;\varepsilon,t) 
&=& 
\whmfu(\vp,\vp') + \int d^2\vp''\,\whmfu(\vp,\vp'')\,
\whmfGra(\vp'',\vr;\varepsilon,t)\circ\whmftra(\vp'',\vp',\vr;\varepsilon,t)
\,,
\eer
\ber\label{eq-T-matrix_K}
\whmftk(\vp,\vp',\vr;\varepsilon,t) 
&=& 
\int d^2\vp''\,
\whmftr(\vp,\vp'',\vr;\varepsilon,t)
\circ
\whmfGk(\vp'',\vr;\varepsilon,t)\circ\whmfta(\vp'',\vp',\vr;\varepsilon,t)
\,.
\eer
The impurity T-matrices are determined by the electron-impurity vertex, $\whmfu(\vp,\vp')$, and the quasiclassical propagators. To leading order in $\sml$ the corresponding self energies also depend only on the mean impurity density, $n_{\text{imp}}$. Multiple scattering by more than one impurity, quantum interference and coherent backscattering of conduction electrons are higher order in $\sml$.~\cite{ram86}

The non-equilibrium quasiclassical equations, normalization conditions, and self-consistency equations represented by the self energy diagrams in Fig.\ref{fig-Sigma-small} form a set of nonlinear integro-differential equations that determine the non-equilibrium quasiparticle distribution functions, the dynamics of the order parameter and the non-equilibrium charge and current response. These equations are accurate to leading order in the expansion parameters of Fermi-liquid theory (e.g. $\sml = \{\kb T/E_f\,, \hbar/p_f\xi_0\,,\hbar/p_f \ell\,,\hbar/\tau E_f\,,\hbar\omega/E_f\,,\ldots$). In the following I discuss the effects of disorder on charge transport in normal and superconducting alloys.

\section{Charge Diffusion}\label{sec-Normal_State-Charge_Diffusion}

For the normal Fermi-liquid state the quasiclassical equations reduce to the Landau-Boltzmann transport equation, which can be expressed as a transport equation for the Keldysh propagator. For the particle sector this equation is given by
\ber\label{eq-Keldysh_Transport_normal-state}
i\left(\frac{\partial \mfgk}{\partial t} + \vv_{\vp}\cdot\gradr\mfgk\right) 
- 
\left[v + \mfsig_{\mbox{\tiny mf}}\,,\,\mfgk\right]_{\circ}
=2\pi i\,\mfsigk+\left\{\mfsig_{\mbox{\tiny c}}\,,\,\mfgk\right\}_{\circ}
\,.
\eer
The left side of Eq. \ref{eq-Keldysh_Transport_normal-state} describes the smooth evolution of the quasiparticle distribution function in phase space under the action of external fields, $v(\vp,\vr;t)$, and the molecular field self energy in Eq.~\eqref{eq-molecular_field}, 

The terms contributing to the collision integral are grouped together on the right side of Eq. \ref{eq-Keldysh_Transport_normal-state}, and are proportional to $\mfsigk$ and $\mfsigc\equiv\left(\mfsigr-\mfsiga\right)$. In the example to follow I evaluate the impurity self energies in the second-order Born approximation, i.e. the first- and second-order diagrams shown in Fig. \ref{fig-Sigma-small}(e), 
\begin{equation}\label{eq-Sigma_Born}
\mfsigrak(\vp,\vr;\varepsilon,t)=\int\,d^2\vp'\, w(\vp,\vp')\,\mfgrak(\vp',\vr;\varepsilon,t)\,,
\end{equation}
where $w(\vp,\vp')=n_{\mbox{\tiny imp}} N_f |\mfu(\vp,\vp')|^2$ is the impurity-scattering rate for quasiparticles with momentum $\vp$ scattering to a final state with momentum $\vp'$ on the Fermi surface in the Born approximation.\footnote{Note that the first-order term $\mfu(\vp,\vp)$ drops out of the normal-state transport equation.}

Quasiparticle transport results from the coupling to electromagnetic field represented by the vector and scalar potentials, $v=\frac{e}{c}\vv_{\vp}\cdot\vA(\vr,t)-e\Phi(\vr,t)$. For weak fields I obtain the \emph{linearized} transport equation,
\begin{equation}\label{eq-linearized_transport-equation}
\hspace*{-3mm}
\frac{\partial \mfgk}{\partial t}\ns +\ns \vv_{\vp}\cdot\gradr\mfgk
+\left(\frac{\partial \mfgk_{\mbox{\tiny eq}}}{\partial\varepsilon}\right)
\pder{}{t}\left[\frac{e}{c}\vv_{\vp}\cdot\vA-e\Phi\right]
=I[\mfgk]
\,.
\end{equation}
where the collision integral on the right side of Eq. \ref{eq-linearized_transport-equation} reduces to
\begin{equation}\label{eq-impurity_collision_integral}
\hspace*{-3mm}
I[\mfgk]\ns=\ns\int\,d^2\vp'\,w(\vp,\vp')\left[\mfgk(\vp',\vr;\varepsilon,t) - \mfgk(\vp,\vr;\varepsilon,t)\right]
\,.
\end{equation}
The collision integral describes the net change in the distribution function for quasiparticles of momentum $\vp$ resulting from the ``scattering out'' of and ``scattering in'' to the state $\vp$. For simplicity I assume the quasiparticle-impurity scattering rate is dominated by scattering in the s-wave channel, in which case $w(\vp,\vp')\approx 1/\tau$, independent of the initial and final momenta on the Fermi surface. Solutions to the transport equation for the linear response function $\delta\mfgk(\vp,\vr;\varepsilon,t) = \mfgk(\vp,\vr;\varepsilon,t)-\mfgk_{\text{eq}}(\vp;\varepsilon)$ determine the charge and current density response to electro-magnetic fields.
The linearized transport equation can be Fourier transformed to obtain the linear response function $\delta\mfgk(\vp,\vq;\varepsilon,\omega)$. The Fourier amplitudes for the EM field are $\vE(\vq,\omega)=\frac{i\omega}{c}\vA(\vq,\omega)-i\vq\Phi(\vq,\omega)$. The charge density response is given by
\begin{equation}
\delta n(\vq,\omega)=e\,N_f\,\left[\langle\delta\mfgk\rangle(\vq,\omega)-e\Phi(\vq,\omega)\right]
\,,
\end{equation}
where $\langle\delta\mfgk\rangle\equiv\int\,d^2\vp\,\int\frac{d\varepsilon}{4\pi i}\delta\mfgk$ is the Fermi-surface average of the equal-time Keldysh distribution function. 

The solution to the linearized transport equation for $\langle\delta\mfgk\rangle$  is given by 
\ber
\langle\delta\mfgk\rangle
=
\frac{-i\omega}{1-\chi(\vq,\omega)}
\Big(
-
e\Phi(\vq,\omega)\,\tau\,\chi(\vq,\omega)
+
\frac{e}{c}\,\vA(\vq,\omega)\cdot\pmb{\chi}
\Big)
\,,
\eer
where the scalar and vector response functions are defined by
\ber
\chi
&=&
\int\,d^2\vp\frac{1}{1-i\omega\tau+i\vq\cdot\vell_{\vp}}
\,,
\\
\pmb{\chi}
&=&
\int\,d^2\vp
\frac{\vell_{\vp}}
{1-i\omega\tau+i\vq\cdot\vell_{\vp}}
\,,
\eer
and $\vell_{\vp}=\vv_{\vp}\tau$ is the mean free path for quasiparticles propagating along the classical trajectory defined by the Fermi velocity $\vv_{\vp}$. 
For long-wavelength ($q\ell\ll 1$), low-frequency ($\omega\tau\ll 1$) EM fields I expand the scalar and vector response functions to leading order in $\omega\tau$ and $\vq\cdot\vell_{\vp}$. In this limit the charge density is governed by a diffusion pole of the charge response function,
\begin{equation}
\delta n(\vq,\omega)=-N_f e^2\,
\left[\frac{\vq\cdot\pmb{\cal D}\cdot\vq}
{-i\omega + \vq\cdot\pmb{\cal D}\cdot\vq}
\right]
\,\Phi(\vq,\omega)
\,.
\end{equation}
Equivalently, the Fermi-surface averaged Keldysh propagator obeys an inhomogeneous diffusion equation,
\begin{equation}
\left(\pder{}{t} -\nabla_i{\cD}_{ij}\nabla_j\right)\langle\delta\mfgk\rangle(\vr,t) 
= e\pder{\Phi}{t}
\,,
\end{equation}
where
\begin{equation}\label{eq-Dtensor}
\cD_{ij} = \tau\,\int\,d^2\vp\,(\vv_{\vp})_i(\vv_{\vp})_j
\,,
\end{equation}
is the diffusion tensor, reflecting the anisotropy of the Fermi surface and Fermi velocity. For an isotropic Fermi surface I obtain the standard result, $\cD_{ij}=\nicefrac{1}{3}v_f^2\tau\delta_{ij}$. 

A similar analysis of the charge current response defines the conductivity tensor,
\be
\vj(\vq,\omega)
=
N_f\,\langle e\vv_{\vp}\,\delta\mfgk\rangle(\vq,\omega) 
= 
\pmb{\sigma}(\vq,\omega)\cdot\vE(\vq,\omega)
\,.
\ee
For $q\rightarrow 0$ the a.c. conductivity reduces to Drude's result,
\be
\pmb{\sigma}(\omega)=\frac{N_f e^2\pmb{\cD}}{1-i\omega\tau}
\,,
\ee
and Einstein's relation, $\pmb{\sigma}(0)=e^2N_f\,\pmb{\cD}$, relating the 
d.c. conductivity and diffusion tensors.~\cite{smith89}

\section{Usadel's Equations}\label{sec-Usadel_Equations}

The timescale for Cooper pair formation in the superconducting state is $t_{coh}=\hbar/2\pi k_BT_c$. The corresponding spatial scale for the pairing corrleations in the clean limit is then given by $\xi_0=v_f\,t_{coh}$ for ballistic electrons moving with the Fermi velocity.
In a disordered metal the ballistic estimate for the pair correlation length breaks down if $\ell\lesssim\xi_0$. Thus, for ``dirty'' superconductors the correlation length is reduced by the diffusive motion of the quasiparticles. The pairing correlation length now becomes, $\xi^2=3{\cal D}t_{coh}$, or
\be
\xi=\sqrt{\frac{3\hbar\cD}{2\pi k_B T_c}}=\sqrt{\ell\xi_0}
\,.
\ee
In the limit $\ell\ll\xi_0$ Usadel obtained a simplified set of diffusion-type equations for the quasiclassical propagators averaged over the Fermi surface.\cite{usa70} Diffusive motion of quasiparticles in disordered metals has dramatic effects on the transport properties of disordered superconductors.~\cite{sch75,ale85,ram86,kamenev11} Below I derive the Usadel equations for the Keldysh formulation of nonequilbrium quasiclassical theory following the method developed in Ref.~\onlinecite{ale85}.

\subsection{Expansion in $\ell/\xi$}

A key feature of the ``dirty limit'' is that the impurity self-energy is the dominant term compared to all other terms in the transport equations. If I assign $\sm\equiv\ell/\xi\ll 1$, then the dominant contribution to the self-energy is of order 
\begin{equation}
[\Sigma_{\mbox{\tiny imp}}]=\frac{\xi_0}{\ell}={\sm}^{-2}
\,,
\end{equation}
relative to the superconducting energy scale, $2\pi T_c$.\footnote{Central to this estimate is that $T_c$ is nearly insensitive to non-magnetic disorder for s-wave superconductors.\cite{and59}} Similarly, other terms in the transport equation are of order,
\ber
[\hbar\vv_{\vp}\cdot\gradr]
&=&
\sm^{-1} 
\,,
\\
{[\varepsilon\whtauz-\whDelta]}
&=&
\sm^0
\,.
\eer
The quasiparticle coupling to a magnetic field via the vector potential is of order 
\begin{equation}
[\frac{e}{c}\vv_{\vp}\cdot\vA]\approx\frac{v_f}{2\pi T_c}\frac{e}{c}\,H_c\lambda
\approx
\frac{\xi_0}{\xi}
\approx
\sqrt{\frac{\xi_0}{\ell}}
=
\sm^{-1}
\,.
\end{equation}

The expansion in powers of $\sm$ is used to reduce the quasiclassical transport equations to the Usadel diffusion-type equations. The procedure is to formally expand the propagators in powers of $\sm$,
\begin{equation}
\whmfGrak = \whmfGrak_0  + \whmfGrak_1 + \whmfGrak_2 + \ldots
\,,
\end{equation}
with $[\whmfGrak_n]=\sm^n$. Then separate the transport and normalization equations into terms corresponding to powers in $\sm$. The leading order terms contributing to the transport equations in powers of $\sm$ are,
\ber
&
\hspace*{-5mm}
\sm^{-2}:
&
\left[\whSigra_{\text{imp}}\,,\,\whmfGra_0\right]_{\circ}=0
\label{RA-small-2}
\\
&
\hspace*{-5mm}
\sm^{-1}:
&
i\hbar\vv_{\vp}\cdot\gaugegrad\whmfGra_0 - \left[\whSigra_{\mbox{\tiny imp}}\,,\,\whmfGra_1\right]_{\circ}=0
\,,
\label{RA-small-1}
\\
&
\hspace*{-5mm}
\sm^{0}:
&
\left[\varepsilon\whtauz-\whDelta\,,\,\whmfGra_0\right]_{\circ}
+
i\hbar\vv_{\vp}\cdot\gaugegrad\whmfGra_1
-\left[\whSigra_{\mbox{\tiny imp}}\,,\,\whmfGra_2\right]_{\circ}=0
\,,
\label{RA-small-0}
\eer
where
\be\label{eq-gaugegrad}
\gaugegrad\widehat{X}\equiv\gradr\widehat{X} - i\frac{e}{c}\,\left[\vA\,\whtauz\,,\,\widehat{X}\right]_{\circ}
\,,
\ee
is the covariant derivative of $\hat{X}$ with respect to local gauge transformations. 
These equations are supplemented by the expansion of the normalization condition in powers of $\sm$:
\ber
&
\hspace*{-7mm}
\sm^{0}:
&
\whmfGra_0\circ\whmfGra_0 = -\pi^2\tone 
\,,
\label{RA-norm0}
\\
&
\hspace*{-7mm}
\sm^{1}:
&
\whmfGra_0\circ\whmfGra_1
+
\whmfGra_1\circ\whmfGra_0
=0
\,,
\label{RA-norm1}
\\
&
\hspace*{-7mm}
\sm^{2}:
&
\whmfGra_0\circ\whmfGra_2
+
\whmfGra_2\circ\whmfGra_0
+
\whmfGra_1\circ\whmfGra_1 = 0
\,.
\label{RA-norm2}
\eer

The impurity self-energy is isotropic in momentum space, and to leading order is proportional to the Fermi-surface average of the propagator, 
\be\label{Sigma-imp}
\whSigra_{\mbox{\tiny imp}} = \frac{1}{\tau} \langle\whmfGra(\vp,\vr;\varepsilon,t)\rangle_{\vp}
\,.
\ee
Equation \ref{RA-small-2}, combined with the normalization conditions (Eqs. \ref{RA-norm0}-\ref{RA-norm2}), imply that $\whmfGra_0(\vr;\varepsilon,t) = \langle\whmfGra(\vp,\vr;\varepsilon,t)\rangle_{\vp}$, and the higher-order angular averages vanish, $\langle\whmfGra_{n}(\vp,\vr;\varepsilon,t)\rangle_{\vp} = 0$, where $\langle\ldots\rangle_{\vp}\equiv\int d^2\vp (\ldots)$ is the angular average over the Fermi surface.
Equations \ref{RA-small-1}, \ref{RA-norm0} and \ref{RA-norm1} can be used to obtain the first-order correction to the propagator, 
\begin{equation}\label{gRA-1}
\whmfGra_1(\vp,\vr;\varepsilon,t) =-\frac{1}{\pi}\,\tau\vv_{\vp}\cdot
\left(
\whmfGra_0\circ\gaugegrad\whmfGra_0
\right)
\,.
\end{equation}
Usadel's equation for $\whmfGra_0$ is then obtained by inserting Eq. \ref{gRA-1} for $\whmfGra_1$ into Eq. \ref{RA-small-0} and integrating all terms in the resulting transport equation over the Fermi surface,
\begin{equation}\label{eq-Usadel_RA}
\frac{1}{\pi}\gaugegrad\cdot\pmb{\cD}\cdot\left(\whmfGra_0\circ\gaugegrad\whmfGra_0\right) 
\ns+\ns 
\left[\varepsilon\hat{\tau}_3 - \whDelta\,,\whmfGra_0\right]_{\circ} 
= 0
\,,
\end{equation}
where $\pmb{\cD}$ is the normal-state diffusion tensor given by Eq.~\ref{eq-Dtensor}. Note that the term involving the commutator of the impurity self-energy with the second-order correction to the propagator, $\whmfGra_2$, drops out of the transport equation upon averaging over the Fermi surface since $\langle\whmfGra_n\rangle_{\vp}=0$ for $n\ge 1$.

A similar analysis carried out for $\whmfGk_0$ yields,
\begin{eqnarray}\label{eq-Usadel_K}
\frac{1}{\pi}\gaugegrad\cdot\pmb{\cD}\cdot\left(\whmfGr_0\circ\gaugegrad\whmfGk_0\right) 
+
\frac{1}{\pi}\gaugegrad\cdot\pmb{\cD}\cdot\left(\whmfGk_0\circ\gaugegrad\whmfGa_0\right) 
+
\left[\varepsilon\hat{\tau}_3 - \whDelta\,,\whmfGk_0\right]_{\circ} 
= 0
\,,
\end{eqnarray}
\be\label{eq-QC_normK0}
\hspace*{-20mm}
\mbox{and}\qquad\whmfGr_0\circ\whmfGk_0 + \whmfGk\circ\whmfGa_0 = 0 
\,.
\ee
The latter condition is satisfied by introducing the Nambu matrix distribution function, $\widehat{\mathsf\Phi}$,
\be\label{eq-Nambu_distribution_function}
\whmfGk_0 = \whmfGr_0\circ\widehat{\mathsf\Phi} -\widehat{\mathsf\Phi}\circ\whmfGa_0 
\,.
\ee
In the low-frequency, long-wavelength limit, $\omega\ll \Delta$, $q v_f\ll\Delta$, Shelankov's projection operators for the local particle and hole sectors can be used to derive transport equations for Bogoliubov quasiparticles in the diffusive limit.\cite{she80} See also Ref.~\onlinecite{kamenev11} for applications of the Keldysh-Usadel theory to collective modes in disordered superconductors.
For a detailed derivation of the kinetic equations and interpretation of $\widehat{\mathsf\Phi}$ in the weak scattering (ballistic) limit see Ref.~\onlinecite{ser83}.

\subsection{Equilibrium Usadel Equations}\label{sec-Equilibrium_Usadel}

The first observation on the structure of the equations for $\whmfGrak_0$ is that, in contrast to Eilenberger's transport equation, Usadel's equations are second-order differential equation of the diffusion type. 
For inhomogeneous equilibrium states the energy-time convolution operator drops out and Eq. \ref{eq-Usadel_RA} can be analytically continued ($\varepsilon\rightarrow i\varepsilon_n$) to obtain Usadel's equation for the Matsubara propagator,
\begin{equation}\label{eq-Usadel_M}
\frac{1}{\pi}\gaugegrad\cdot\pmb{\cD}\cdot\left(\whmfG_0\,\gaugegrad\,\whmfG_0\right) 
\ns+\ns 
\left[i\varepsilon_n\hat{\tau}_3 - \whDelta\,,\whmfG_0\right]
= 0
\,,
\end{equation}
which also obeys the Eilenberger normalization condition, 
\be\label{eq-Normalization_M}
\left[\whmfG_0(\vr;\varepsilon_n)\right]^2 = -\pi^2\,\tone
\,.
\ee

\subsection{Anderson's Theorem}\label{sec-Anderson_Theorem}

The second observation is that the impurity self energy has dissappeared. Disorder appears in Usadel's equation only via the normal-state diffusion coefficient multiplying the second-order derivative term. Thus, for a dirty superconductor with a homogeneous order parameter in the absence of magnetic fields the solution of Eqs. \ref{eq-Usadel_M} and \ref{eq-Normalization_M} for the equilibrium propagator is \emph{independent of disorder} and given by 
\be\label{eq-Propagator_M}
\whmfG_0(\varepsilon_n) = -\pi\frac{i\varepsilon_n\whtauz - \whDelta}{\sqrt{\varepsilon_n^2 + |\Delta|^2}}
\,.
\ee
The implications of this result are that the bulk thermodynamic properties in zero field are insensitive to disorder. In the weak-coupling limit the mean-field pairing self energy is determined by Eq. \ref{eq-Delta}. From Eq. \ref{eq-Propagator_M}, and analytic continuation to Matsubara energies, the gap equation becomes
\be
\frac{1}{g} = \pi T\sum_{n}^{|\varepsilon_n| \le \varepsilon_c} 
                         \frac{1}{\sqrt{\varepsilon_n^2 + |\Delta|^2}} 
\,.
\ee
The logarithmic dependence on the cutoff is regularized by subtracting the kernel of the linearized gap equation that determines $T_c$ to express the gap equation as a convergent Matsubara sum, as well as to remove the cutoff and pairing interaction in favor of $T_c$,
\be
\ln(T/T_c) = \pi T\sum_{n=-\infty}^{+\infty}
                  \left[\frac{1}{\sqrt{\varepsilon_n^2 + |\Delta|^2}}  - \frac{1}{|\varepsilon_n|}\right]
\,.
\ee
Similarly, the quasiparticle density of states obtatined from analytic continutation of Eq. \ref{eq-Propagator_M} to the real energy axis, $i\varepsilon_n\rightarrow \varepsilon + i0^{+}$, and the spectral function defined by Eq. \ref{eq-LDOS} reduces to,
\begin{equation}
N(\varepsilon)=N_f\,\frac{|\varepsilon|}{\sqrt{\varepsilon^2-|\Delta|^2}}\,\Theta(\varepsilon^2-|\Delta|^2)
\,.
\end{equation}
Thus, the superconducting transition temperature, $T_c$, gap amplitude, $\Delta(T)$, excitation spectrum and thermodynamic potential are insensitive to disorder for isotropic s-wave pairing, what is often referred to as Anderson's ``theorem''.\cite{and59}$^{'}$\footnote{For an extension of the Anderson theorem to a specific class of anisotropic unconventional superconductors with anisotropic impurities See Ref.~\onlinecite{fom18}.}
These results were derived independently by Abrikosov and Gorkov\cite{abr59b} based on a field-theoretical formulation of the pairing theory and the impurity averaging technique of Edwards,\cite{edw58} the latter of which generates the T-matrix theory for the electron-impurity self energy defined by Eqs. \ref{eq-Sigma_imp}-\ref{eq-T-matrix_K}.
Note that for pairing that is isotropic on the Fermi surface the insensitivity of the gap and $T_c$ does not depend on the strong disorder limit, i.e. $\hbar/\tau \gg 2\pi T_c$. However, conventional superconductors in the clean limit typically have anisotropic pairing interactions even in the $\point{A}{1g}$ pairing channel, and as a result impurity scattering that samples different gap amplitudes on the Fermi surface leads to Andreev scattering, pair breaking, and for weak impurity scattering rates, $\hbar/2\pi\tau T_{c_0} \lesssim 1$, to a weak suppression of $T_c$,\cite{mar63,hoh64,zar22}
which persists until the impurity scattering rate is sufficiently fast to average the anisotropic pairing interaction over the bandwidth of attraction, $1/\tau\gg\omega_c$.\cite{rai86}

\subsection{Current Response}\label{sec-Current_Response}

Significant differences between the clean and dirty limit appear for \emph{inhomogeneous} superconducting states, even spatially uniform current-carrying states. Persistent currents, i.e. equilibrium supercurrents, are described by a spatially varying phase of the order parameter. The order parameter for spin-singlet, s-wave pairing can be expressed in terms of the amplitude, $|\Delta(\vr)|$ and phase, $\vartheta(\vr)$,  
\be\label{eq-OP-inhomogeneous}
\whDelta(\vr)=|\Delta(\vr)|(i\sigma_y)\left[e^{+i\vartheta(\vr)}\tp+e^{-i\vartheta(\vr)}\tm\right]
\,,
\ee
where $\widehat{\tau}_{\pm} = (\tx \pm i \ty)/2$ combined with $\tz$ represent the circular basis for the Nambu matrices (see Appendix \ref{appendix-Nambu_Algebra}). In this basis the matrix propagator can be expressed as 
\be\label{eq-propagator-inhomogeneous}
\whmfG_0(\vr;\varepsilon_n)\ns=\ns\mfg(\vr;\varepsilon_n)\tz 
\ns+\ns i\sigma_y\left[\mff(\vr;\varepsilon_n)\tp\ns+\ns\mfuf(\vr;\varepsilon_n)\tm\right]
\,.
\ee  

Given the zeroth-order Matsubara propagator that satisfies Eqs. \ref{eq-Usadel_M} and \ref{eq-Normalization_M} for a spatially varying order parameter, or ``pair potential'', the leading-order correction to the propgator can be calculated from the analytic continuation of Eq.~\ref{gRA-1},
\be
\whmfG_1(\vp,\vr;\varepsilon_n) =\frac{\tau}{i\pi}\,\left[\whmfG_0\,\vv_{\vp}\cdot\gaugegrad\,\whmfG_0\right]
\,,
\ee
which yields Usadel's result for the supercurrent in the dirty limit,~\cite{usa70}
\ber\label{eq-Current_Matsubara} 
\vj(\vr) 
=
N_f\,T\sum_{\varepsilon_n}\,\int d^2\vp\,(e\vv_{\vp})\,
\nicefrac{1}{2}\Tr{\tz\whmfG_{1}(\vp,\vr;\varepsilon_n)}
=
-
\frac{eN_f\cD}{\pi}\,T\sum_{\varepsilon_n}\,
\left[
\mfuf\,\frac{\hbar}{i}\,\gaugegrad\mff
- 
\mff\,\frac{\hbar}{i}\,\gaugegraddag\mfuf 
\right]
\,,
\eer
where the covariant derivative reduces to $\gaugegrad = \gradr-i\frac{2e}{\hbar c}\vA$.

The dependence on the Matsubara energy encodes both the spectral resolution of the Fermionic states contributing to macroscopically occupied Cooper pair condensate \emph{and} the reduction of the current due to thermal excitations out of the condensate. These features of the current are revealed by transforming to the real energy axis (c.f. Appendix~\ref{appendix-Matsubara_Transformation}),

\be\label{eq-Current_Spectral-Representation} 
\vj(\vr) 
=
-
\frac{eN_f\cD}{\pi}\,\int_{-\infty}^{+\infty}\ns\ns d\varepsilon\,f(\varepsilon)\,
\left[-\frac{1}{\pi}\Im
\Bigg\{
\mfufr(\vr;\varepsilon)\,\frac{\hbar}{i}\,\gaugegrad\mffr(\vr;\varepsilon)
-
\mffr(\vr;\varepsilon)\,\frac{\hbar}{i}\,\gaugegraddag\mfufr(\vr;\varepsilon)
\Bigg\}
\right]
\,,
\ee
where $f(\varepsilon)$ is the Fermi distribution. This result has the wave-mechanical structure for the current of charge $2e$ Cooper pairs, but spectrally resolved in terms of the Fermionic states that contribute to the current carried by the Cooper pair condensate. At $T=0$ the negative energy states which comprise the Cooper pair condensate determine the maximum current density, with thermally excited particles and holes leading to a reduction of the supercurrent governed by the Fermi distribution.

For slow spatial variations of the phase on the scale of the coherence length, $\xi|\grad\vartheta|\ll 1$, the currents are small in magnitude compared to the maximum sustainable supercurrent, i.e. the bulk critical current, and the magnitude of the order parameter is to good approximation constant and given by the bulk equilibrium value. In this London limit the order parameter is given by, $\Delta(\vr)=|\Delta|\,\exp{i\vartheta(\vr)}$, and the corresponding anomalous propgator is given by
\ber
\mffr &=& \frac{\pi|\Delta|e^{+i\vartheta(\vr)}}{\sqrt{|\Delta|^2 - (\varepsilon + i 0^{+})^2}}
\,,
\eer
and $\mfufr(\vr;\varepsilon) = \mffr(\vr;-\varepsilon)^*$. The resulting spectral current density becomes,
\ber
-\frac{1}{\pi}\Im
\Bigg\{
\mfufr(\vr;\varepsilon)\,\frac{\hbar}{i}\,\gaugegrad\mffr(\vr;\varepsilon)
-
\mffr(\vr;\varepsilon)\,\frac{\hbar}{i}\,\gaugegraddag\mfufr(\vr;\varepsilon)
\Bigg\}
&=&
\left[-\frac{1}{\pi}\Im
\frac{4\pi^2|\Delta|^2}{(\varepsilon+i0^+)^2 - |\Delta|^2}
\right]
\,\vp_s
\nonumber
\\
&=&
2\pi^2|\Delta|\,
\Big[
\delta(\varepsilon-|\Delta|) - \delta(\varepsilon+|\Delta|)
\Big]
\,\vp_s
\,,
\eer
\be\label{eq-condensate_momentum}
\hspace*{-22mm}\mbox{where}\qquad\qquad
\vp_s=\onehalf\left(\hbar\gradr\vartheta-\frac{2e}{c}\vA\right)
\,,
\ee 
is the gauge-invariant momentum per particle of the condensate. Note that the structure of the spectral current density in the dirty limit leads to a cancellation of the contribution from states represented by the branch cuts, leaving only the contributions from isolated poles at $\varepsilon_{\pm} = \pm |\Delta|$.
This result is in sharp contrast to the spectral current density in the clean limit where a branch cut ensures that the entire negative energy spectrum with $\varepsilon\le -|\Delta|$ contributes to the $T=0$ supercurrent.
My interpretaion for only the states at $\varepsilon=\pm|\Delta|$ contributing to the spectral current density is that for any realization of a random distribution of impurities Tomasch oscillations~\cite{tom65,buc95b} induced in the spectral current density due to impurity-induced Andreev scattering by the spatially varying phase destructively interfere for all energies except $\varepsilon\rightarrow\pm|\Delta|$ where the Tomash wavelength $\Lambda_{\text{T}}(\varepsilon) = \hbar v_f/\sqrt{\varepsilon^2-|\Delta|^2}$ diverges. 

\begin{figure}[t]
\includegraphics[width=0.75\columnwidth]{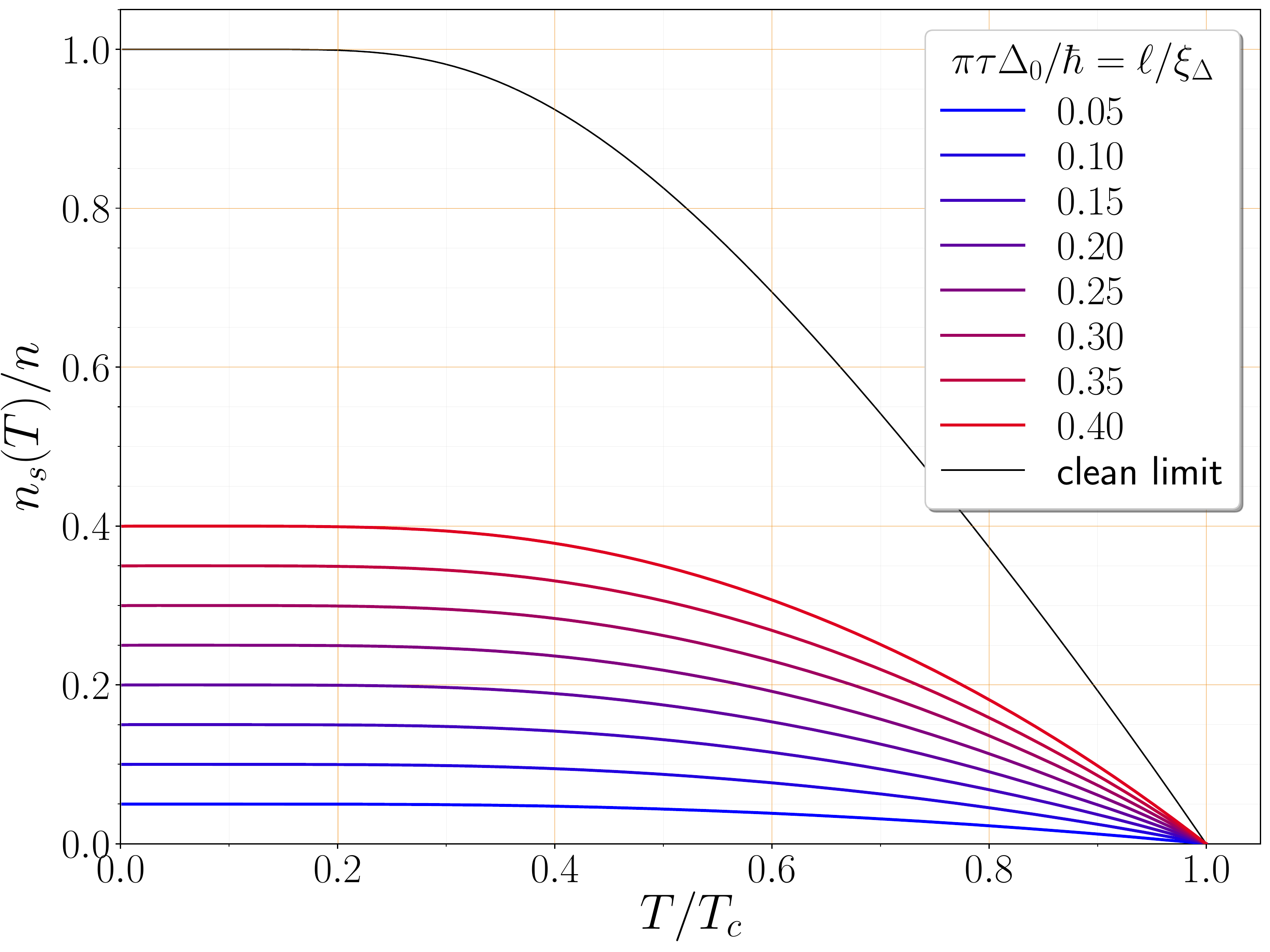}
\caption{
The superfluid fraction, $n_s(T)/n$, determines the London penetration depth in the low-field limit.
}
\label{fig-Superfluid_Fraction}
\end{figure}

Thus, in the dirty limit the resulting supercurrent obtained from Eq. \ref{eq-Current_Spectral-Representation} becomes,
\ber
\vj(\vr) 
= 2\pi\,eN_f\cD\,|\Delta|\,\tanh\left(\frac{|\Delta|}{2T}\right)\,\vp_s
\equiv e\,n_s(T)\,\vv_s 
\,,
\label{eq-Current_London-Limit}
\eer 
and thus a superfluid fraction given by
\be\label{eq-ns}
n_s = n \left(\frac{\ell}{\xi_{\text{$\Delta$}}}\right)\,
\left(\frac{|\Delta|}{\Delta_0}\right)
        \tanh\left(\frac{|\Delta|}{2T}\right) 
\,,
\ee
where $\Delta_0=|\Delta(0)|$, $n=\nicefrac{2}{3}N_f v_f p_f$ is the total electron density comprising the Fermi sea, and $\xi_{\text{$\Delta$}} = \hbar v_f/\pi\Delta_0$ is the zero-temperature coherence length in the clean limit. Note that I have introduced the superfluid velocity field $\vv_s\equiv\vp_s/m^*$, where the quasiparticle effective mass is defined by $m^*\equiv p_f/v_f$. Thus, although $T_c$ and the amplitude, $\Delta$, of the condensate order parameter are insensitive to disorder, the current that can be transported by the condensate is dramatically reduced in the dirty limit, $\ell/\xi_{\text{$\Delta$}}\ll 1$ due to destructive interference for states with $\varepsilon<-|\Delta|$.

The superfluid fraction determines the London penetration depth,
\be\label{eq-London_Penetration_Depth}
\frac{1}{\lambda_{\text{L}}^2} = \frac{4\pi n_s\,e^2}{m^*c^2} = 
\frac{1}{\lambda_{\text{L}_0}^2}\,
         \left(\frac{\ell}{\xi_{\text{$\Delta$}}}\right)\, 
         \left(\frac{|\Delta|}{\Delta_0}\right)\,
         \tanh\left(\frac{|\Delta|}{2T}\right)
\,.
\ee 
where $1/\lambda_{\text{L}_0}^2 = 4\pi n e^2/m^*c^2$ determines the zero-temperature pentration length in the clean-limit. Disorder weakens the Meissner screening current, and thus increases field penetration relative to that in the clean limit. Strong disorder also shortens the coherence length so that the Ginzburg-Landau ratio, 
\be
\kappa_{\text{dirty}}\equiv\frac{\lambda_{\text{L}}}{\xi}
\approx
\kappa_{\text{clean}}\,\left(\frac{\xi_{0}}{\ell}\right)
\,, 
\ee
increases with disorder. As a result superconducting alloys in the dirty limit are generally Type II superconductors, even for superconductors that are Type I in the clean limit, i.e. $\kappa_{\text{clean}}=\lambda_{\text{L}_0}/\xi_0 < 1/\sqrt{2}$.\cite{abr59a}

\section{Nonlinear Current Response}\label{sec-Nonlinear_Currents}

The current given by Eq.~\ref{eq-Current_London-Limit} can be extended beyond the linear response limit. The key step is to first ``remove'' the spatially varying phase of the order by a local gauge transformation,
\be
\whDelta(\vr)=\whcU[\vartheta(\vr)]\,\whDelta'\,\whcU^{\dag}[\vartheta(\vr)]
\,,
\ee
where $\whDelta'=|\Delta|\,i\sigma_y\,\tx$ and $\whcU[\vartheta(\vr)]=\exp[i\vartheta(\vr)\tz/2]$ generates a local gauge transformation by phase angle $\vartheta(\vr)$. This representation is used to transform Eq.~\ref{eq-Usadel_M} to the gauge in which the order parameter is real, hereafter the ``real gauge''. The propagator and covariant derivative operator in the new gauge are defined by,
\ber
\whmfG_0^{'}(\vr;\varepsilon_n)
&=&
\whcU^{\dag}[\vartheta(\vr)]\,\whmfG_0(\vr;\varepsilon_n)\,\whcU[\vartheta(\vr)]
\,,
\\
\quad\nonumber
\\
\gaugegrad'\whmfG_0^{'}(\vr;\varepsilon_n)
&=&
\whcU^{\dag}[\vartheta(\vr)]\gaugegrad\,\whmfG_0(\vr;\varepsilon_n)\,\whcU[\vartheta(\vr)]
\,,
\eer
\be
\mbox{where}\quad
\gaugegrad'\widehat{X}=\gradr\widehat{X}-i\left[\vp_s\tz\,,\,\widehat{X}\right]
\,,
\ee
and 
$\vp_s$ is the condensate momentum given by Eq.~\eqref{eq-condensate_momentum}. Usadel's equation in the real gauge is given by Eq. \ref{eq-Usadel_M}, but with $\whmfG_0$ replaced by $\whmfG_0^{'}$ and with $\gaugegrad$ replaced by $\gaugegrad'$.

For the special case of spatially uniform condensate momentum the Usadel equation in the real gauge reduces to 
\begin{equation}\label{eq-Usadel_M-newgauge}
\left[i\varepsilon_n\tz
- 
\frac{2\cD}{\pi}\,p_s^2\,\tz
-
\whDelta'
\,,\whmfG_{0}^{'}\right]
= 0
\,.
\end{equation}
The resulting solution to Eqs. \ref{eq-Usadel_M-newgauge} and \ref{eq-Normalization_M} for the Usadel propagator, transformed back to the original gauge, can now be expressed as
\be\label{eq-Usadel_Propagator-Nonlinear}
\whmfG_{0}(\vr;\varepsilon_n)=-\pi\frac{i\tilde\varepsilon_n\tz-\whDelta(\vr)}{\sqrt{\tilde\varepsilon_n^2+|\Delta|^2}}
\,,
\ee 
with the renormalized Matsubara energies defined by 
\ber
\tilde\varepsilon_n &=& \cZ(\varepsilon_n,p_s)\,\varepsilon_n
\,,
\\
\mbox{with}
\quad
\cZ &\equiv& 1 + 2\cD\,p_s^2\,\frac{\cZ}{\sqrt{\cZ^2\varepsilon_n^2 + |\Delta|^2}}
\,.
\label{eq-Renormalization_Factor}
\eer
Note that $\cZ$ is real for all $\varepsilon_n$, and scales to the limit $\cZ\rightarrow 1$ for $|\varepsilon_n|\rightarrow\infty$.

Equation \ref{eq-Current_Matsubara} for the current response depends on the anomalous propagators,
\be
\mff = \pi\frac{|\Delta|e^{+i\vartheta(\vr)}}{\sqrt{\cZ(\varepsilon_n;p_s)^2\varepsilon_n^2 + |\Delta|^2}}
\,,
\ee 
$\mfuf = \mff^*$, and their covariant derivatives,
\be
\frac{\hbar}{i}\gaugegrad\mff = +2\vp_s\,\mff
\,,\quad\mbox{and}\quad
\frac{\hbar}{i}\gaugegraddag\mfuf = -2\vp_s\,\mfuf
\,.
\ee
The resulting supercurrent is then given by the nonlinear function of the condensate momentum,
\be\label{eq-Current_Nonlinear}
\hspace*{-3mm}
\vj_s=4eN_f\cD\,\left(\pi T\sum_{\varepsilon_n} 
\frac{|\Delta|^2}{\cZ(\varepsilon_n;p_s)^2\,\varepsilon_n^2 + |\Delta|^2}\right)\,\vp_s
\,.
\ee
where $\cZ$ is obtained from the solution of Eq. \ref{eq-Renormalization_Factor} for each Matsubara energy. In the nonlinear regime, the amplitude of the order parameter, $|\Delta|\equiv\Delta(p_s)$, also depends on the condensate momentum and diffusion constant. The gap equation for the self-consistent, mean-field order parameter becomes,
\be\label{eq-Gap_Nonlinear}
\hspace*{-1.5mm}
\ln\frac{T}{T_{c}}
\ns = \pi T \sum_{\varepsilon_n}
\left[
\frac{1}{\sqrt{\cZ(\varepsilon_n;p_s)^2\varepsilon_n^2 + \Delta(p_s)^2}}
\ns-\ns
\frac{1}{|\varepsilon_n|}
\right].
\ee   

\begin{figure}[t]
\includegraphics[width=0.75\columnwidth]{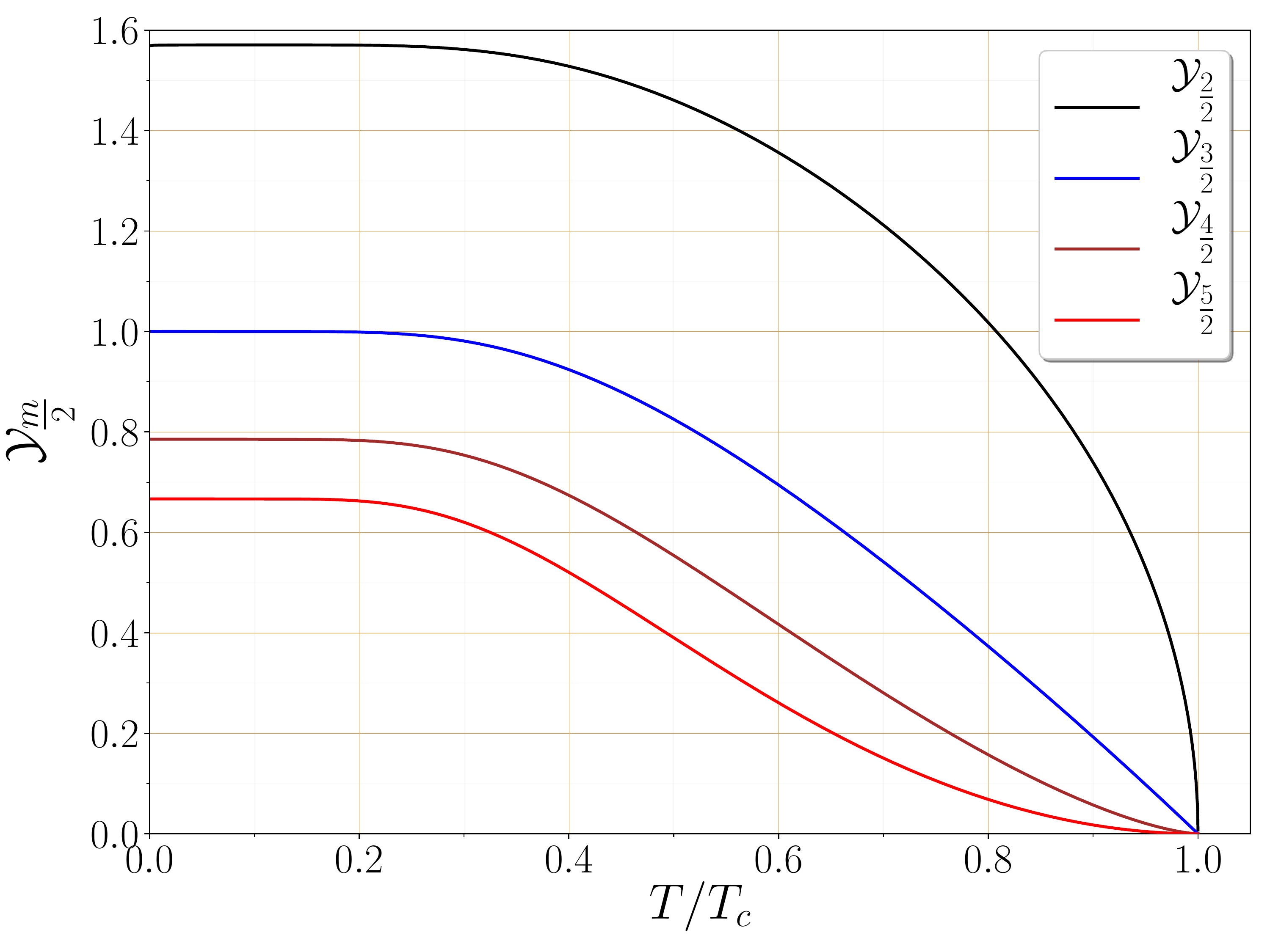}
\caption{
Generalized Yoshida functions, $\y{m}(T)$, that determine gap suppression and the nonlinear Meissner effect.
}
\label{fig-Yoshida_Functions}
\end{figure}

\subsection{Perturbative Nonlinearities}

The leading order nonlinear correction to the current-field equation is obtained by expanding Eqs.~\ref{eq-Current_Nonlinear} and \ref{eq-Gap_Nonlinear} to order $p_s^3$ and $p_s^2$, respectively. The expansion parameter is $2\cD p_s^2\ll|\Delta|$. Setting $\cZ = 1$ on the right side of Eq.~\ref{eq-Renormalization_Factor} gives,
\be\label{eq-Renormalization_Factor-approximate}
\cZ \approx 1 + 2\cD\,p_s^2\,\frac{1}{\sqrt{\varepsilon_n^2 + |\Delta|^2}}
\,.
\ee
As a result the condensate flow field suppresses the gap quadratically in $p_s$. Writing $\Delta(T,p_s)^2=\Delta(T)^2+\Delta^{2}_2(T,p_s)$ and solving the gap equation (Eq. \ref{eq-Gap_Nonlinear}) perturbatively determines the leading order correction to the gap amplitude.
\be\label{eq-Gap_Suppression}
\frac{\Delta^{2}_{2}}{\Delta^2} 
          = -\left(\frac{4\cD\,p_s^2}{\Delta}\right)\,
             \left[\frac{\y{2}(T;\Delta)-\y{4}(T;\Delta)}{\y{3}(T;\Delta)}\right]
\,,
\ee   
where I have introduced the generalized Yoshida functions,
\be\label{eq-y_functions}
\y{m}(T;\Delta) 
\equiv 
\pi T\sum_{\varepsilon_n} 
\frac{\Delta^{m-1}}{\left(\varepsilon_n^2+\Delta^2\right)^{\nicefrac{m}{2}}}
\,,\quad m\ge 2
\,.
\ee

\noindent In terms of the dependence of the gap amplitude on $p_s$,
\ber\label{eq-Gap_Suppression2}
\Delta(T;p_s)
&=&
\Delta(T)\left[1-\beta(T;\tau)\,\left(\frac{p_s}{p_c}\right)^2 \right]
\,,
\\
\hspace*{-5mm}
\beta(T;\tau)
\ns&=&\ns
\frac{2}{3\pi}\left(\pi\tau\Delta_0\right)\,
             \left(\frac{\Delta}{\Delta_0}\right)
             \left[\frac{\y{2}(T)-\y{4}(T)}{\y{3}(T)}\right]
,
\hspace*{7mm}
\eer
where $p_c=\Delta/v_f$, $\Delta_0=\Delta(0)$ is the zero temperature gap for $p_s=0$, $\pi\tau\Delta_0=\ell/\xi_{\mbox{\tiny $\Delta$}}$, and $\y{m}(T) = \y{m}(T;\Delta(T))$. Several relevant Yoshida functions are plotted in Fig.~\ref{fig-Yoshida_Functions}. 

Similarly, the nonlinear correction to the supercurrent can be expressed as 
\be\label{eq-Current_Nonlinear_Perturbative}
\vj_s= e\,n_s(T)\,\left[1-\theta(T;\tau)\left(\frac{p_s}{p_{c}}\right)^2\right]\,\vv_s 
\,,
\ee
and $n_s(T;\tau)$ is the zero-field superfluid fraction,
\be
n_s(T;\tau)=n\,\left(\pi\tau\Delta_0\right)\left(\frac{\Delta(T)}{\Delta_0}\right)\,\frac{2}{\pi}\,\y{2}(T)
\,,
\ee
which is equivalent to Eq.~\ref{eq-ns}, and is plotted in Fig.~\ref{fig-Superfluid_Fraction}. The coefficient of the nonlinear correction is given by
\ber
\theta(T;\tau)=\frac{4}{3\pi}\left(\pi\tau\Delta_0\right)\left(\frac{\Delta(T)}{\Delta_0}\right)
\,
\left[
\left(\y{2}(T)-\y{4}(T)\right)^2\Big/\y{2}(T)\y{3}(T)
+
\left(\y{3}(T) - \y{5}(T)\right)\Big/\y{2}(T)
\right]
\,.
\eer
Note the scales of $n_s/n$, $\theta$ and $\beta$ are determined by $\pi\tau\Delta_0$.

\begin{figure}[t]
\includegraphics[width=0.75\columnwidth]{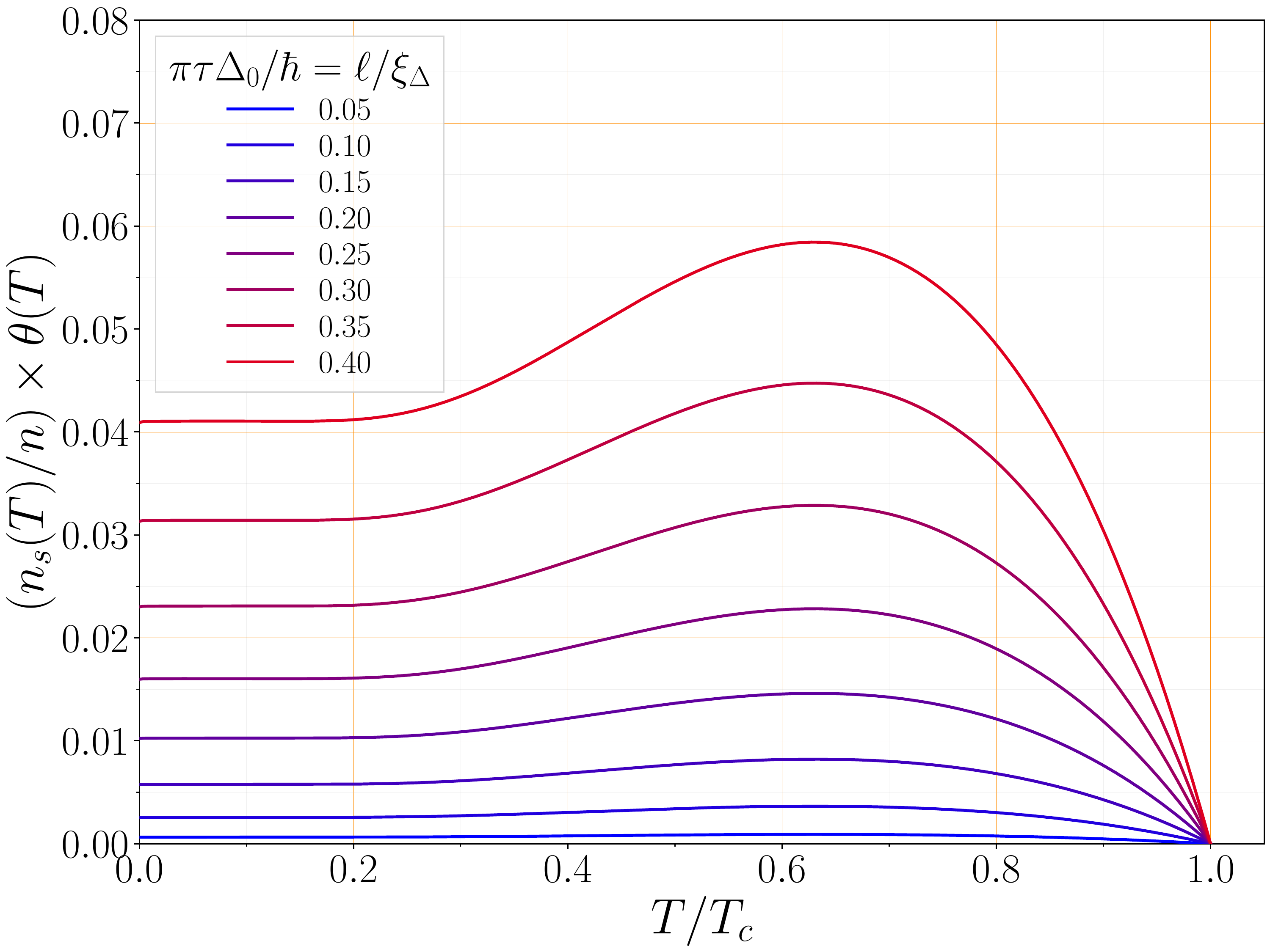}
\caption{
The nonlinear current response is determined by the product $(n_s(T;\tau)/n)\times\theta(T;\tau)$.
}
\label{fig-Nonlinear_Current_Response}
\end{figure}

The zero-temperature limits for the order parameter suppression and nonlinear correction to the current
are obtained by setting $\Delta(0)=\Delta_0\simeq 1.78\,T_c$ and transforming Eq.~\ref{eq-y_functions}
into an integral representation for the Beta function,\cite{abramowitz72}
\ber
\lim_{T\rightarrow 0}\y{m}(T;\Delta) 
=
\onehalf\int_0^1\,dt\,t^{\nicefrac{m-3}{2}}\,(1-t)^{-\nicefrac{1}{2}}
=
\onehalf\cB\left(\nicefrac{m-1}{2},\nicefrac{1}{2}\right)
=
\onehalf\frac{\Gamma\left(\nicefrac{m-1}{2}\right)\,\Gamma\left(\nicefrac{1}{2}\right)}{\Gamma\left(\nicefrac{m}{2}\right)}
\,.
\eer
The resulting zero-temperature limits for the order parameter and current response are given by $n_s(0)= n\left(\pi\tau\Delta_0\right)$, $\beta(0)=\nicefrac{1}{6}\left(\pi\tau\Delta_0\right)$, and $\theta(0)=\nicefrac{4}{3\pi}\left[\nicefrac{\pi}{8}+\nicefrac{2}{3\pi}\right]\left(\pi\tau\Delta_0\right)\approx 0.257\,(\pi\tau\Delta_0)$, which agree with the $T\rightarrow 0$ limits computed numerically and shown in Fig.~\ref{fig-Nonlinear_Current_Response}.

\subsection{Nonlinear Meissner Effect}\label{sec-Nonlinear_Meissner_Effect}

In the Meissner state supercurrents screen an external magnetic field from penetrating into the bulk of the superconductor. 
For weak magnetic fields the screening current is linear in condensate momentum (Eq.~\ref{eq-Current_London-Limit}). Increasing external field drives the current response into the nonlinear regime defined by Eq.~\ref{eq-Current_Nonlinear}. For weak nonlinearity the current response is given by Eq.~\ref{eq-Current_Nonlinear_Perturbative}, and when combined with Amp\`ere's equation and charge conservation, I obtain the nonlinear London equation,
\be
\nabla^2\vp_s - \frac{1}{\lambda_{\text{L}}^2}\,
                \left\{1-\theta(T;\tau)\left(\frac{p_s^2}{p_c^2}\right)\right\}\,\vp_s = 0
\,,
\ee
where $\lambda_{\text{L}}$, defined in Eq.~\ref{eq-London_Penetration_Depth}, is the London penetration depth which defines the field penetration length in the low-field, linear response limit.

\begin{figure}[t]
\includegraphics[width=0.75\columnwidth]{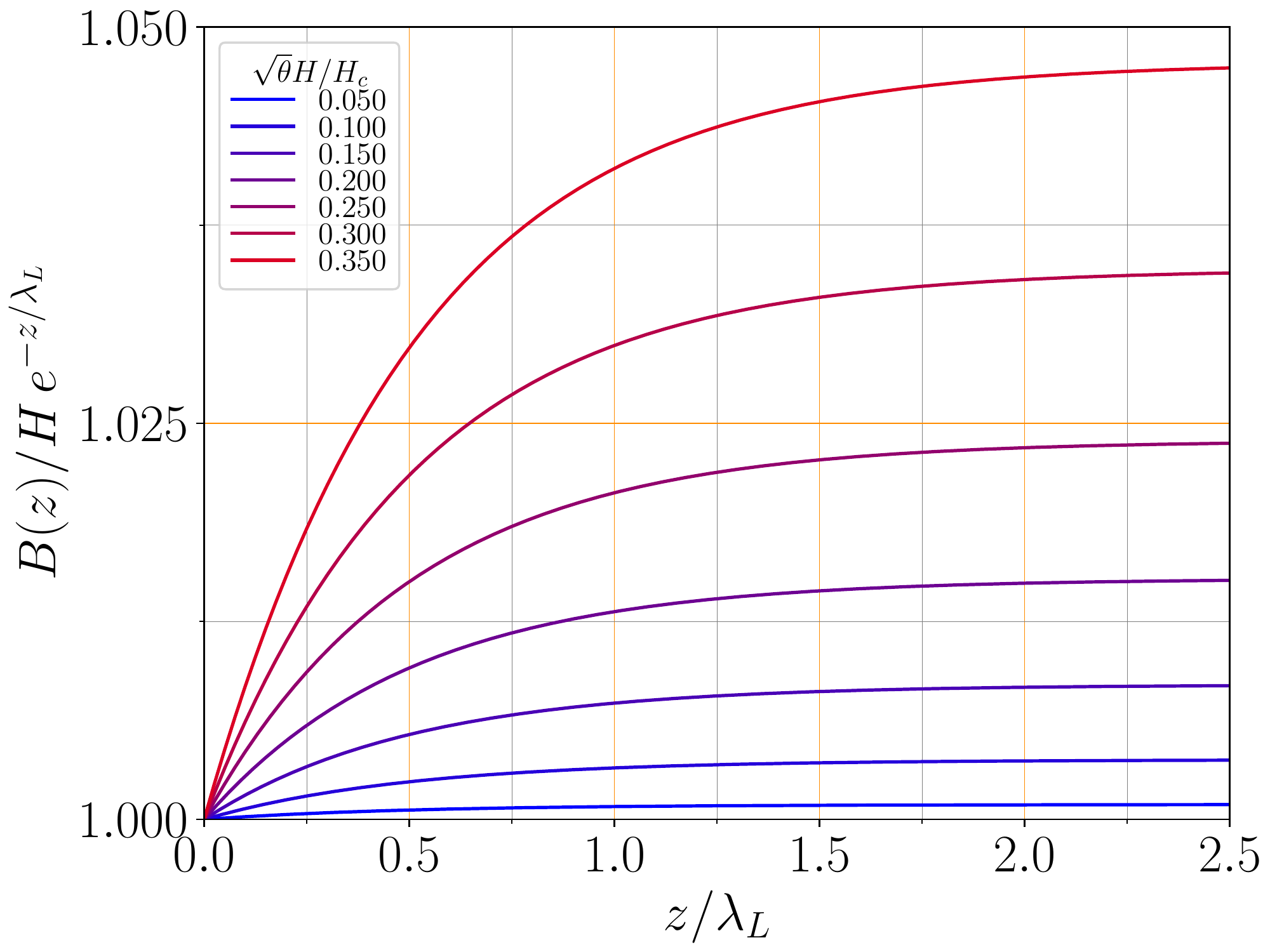}
\caption{
Enhancement of the magnetic field relative to the linear London field, $H\,e^{-z/\lambda_L}$, from the nonlinear correction to the Meissner screening current as a function of depth, $z/\lambda_L$, for a range of scaled external fields, $h=\sqrt{\theta H/H_c}$, from Eq.~\eqref{eq-B_vs_H}.
}
\label{fig-B_vs_z}
\end{figure}

For a planar vacuum-superconducting interface, with $z$ being the distance into the superconductor normal to the interface, and external field at the surface given by $\vH=H\hat\vx$, the screening current and condensate momentum flow parallel to the interface, $\vp_s=p_s(z)\hat\vy$.
The local magnetic field in the superconductor is then given by $\vB=B(z)\hat\vx=\frac{c}{e}\frac{dp_s}{dz}\hat\vx$. Continuity of the field at the vacuum-superconducting interface requires 
\be
\frac{dp_s}{dz}\Bigg\vert_{z=0} = \frac{e}{c}H
\,.
\ee
In addition, the magnetic field, and thus the screening current, also vanish deep inside the superconductor, $\lim_{z\rightarrow\infty}B(z) = 0$. It is convenient to introduce the dimensionless condensate momentum, $u=\sqrt{\theta}p_s/p_c$, and scale distance as, $\zeta=z/\lambda_{\text{L}}$. The differential equation for $u(\zeta)$ becomes,
\be
\frac{d^2u}{d\zeta^2}-u(1-u^2)=0
\,,
\ee
with $b\equiv\der{u}{\zeta}=\sqrt{\theta}\,B(z)/H_c$, supplemented by the boundary conditions,
\be\label{bc}
\frac{du}{d\zeta}\Bigg\vert_{\zeta=0}=\sqrt{\theta}\frac{H}{H_c} \equiv h
\,,\qquad
u({\zeta\rightarrow\infty}) = 0
\,,
\ee
where $H_c=\frac{c}{e}\frac{\Delta}{v_f\lambda_{\text{L}}}$ $=$ $\nicefrac{\Phi_0}{\pi\xi_{\text{$\Delta$}}\lambda_{\text{L}}}$ is the thermodynamic critial field, $\Phi_0=hc/2e$ is the flux quantum and $\xi_{\text{$\Delta$}}=\hbar v_f/\pi\Delta$ is the pair correlation length.
The solution for the magnetic field, 
\be\label{eq-B_vs_H}
b=\frac{h\,e^{2\zeta}}
{\left[h^2/4 + (1-h^2/4)e^{2\zeta}\right]^{3/2}}
\,,
\ee
is derived and discussed in detail in App.~B of Ref.~\onlinecite{xu94a}. Note that if I drop terms of order $h^2$ I recover the linear response solution to the London equation, $B=H\,e^{-z/\lambda_{\text{L}}}$. Increasing the field at the interface suppresses the screening current leading increased field penetration as shown in Fig.~\ref{fig-B_vs_z}.
The penetration depth can be defined in terms of the initial decay rate of the field into the superconductor,
\be
\frac{1}{\lambda(H)}
\ns\equiv\ns
\frac{1}{\lambda_{\text{L}}}
\left[-\frac{1}{h}\frac{db}{d\zeta}\right]_{\zeta=0} 
\ns=\ns
\frac{1}{\lambda_{\text{L}}}
\left[
1\ns-\ns\frac{3}{4}\theta(T;\tau)\left(\frac{H}{H_c}\right)^2
\right]
\,.
\ee
Note that $\lambda(H)$ increases with field as expected, and that this definition of $\lambda(H)$ is equivalent to identifying the penetration depth with the surface reactance, i.e. $\lambda(H)^{-1}\propto j_s(0)/H$. 

\subsection{Microwave Photon Generation}\label{sec-Microwave_Photon_Generation}

The nonlinear screeneing current in Eq.~\eqref{eq-Current_Nonlinear_Perturbative} is valid for low-frequency EM fields, i.e. $\omega\ll 2\Delta$. For Niobium with $T_c\approx 9.3\,\mbox{K}$, $2\Delta(0)/h\simeq 700\,\mbox{GHz}$, in which case for EM fields at GHz frequencies the nonlinear screening current at a vacuum-superconducting interface is to a very good approximation given by Eq.~\eqref{eq-Current_Nonlinear_Perturbative} with $n_s$ and $\theta$ given by their d.c. limits. The dissipative current response from unpaired quasiparticles is exponentially small for $T\ll T_c$.

Thus, the dominant contribution to the current response is given by Eq.~\eqref{eq-Current_Nonlinear_Perturbative}, for any $\vA(\vr,t)$ in the limit $\omega\ll \Delta$. The incident EM field can be strong, i.e. the number of photons in any mode, $N_{k}\gg 1$, but below the threshold for vortex generation. In this limit I can absorb the phase gradient into the vector potential: $\vA+\frac{\hbar c}{2e}\grad\vartheta\rightarrow\vA$ and work in the transverse gauge. The corresponding current response just below the vacuum-superconductor interface can then be expressed as,
\be\label{eq-Current_Nonlinear_A}
\vj_s = -\frac{c}{4\pi\lambda_{\text{L}}^2}
			\left\{1 - \frac{\theta}{A_c^2}\vert\vA\vert^2\right\}\,\vA
\,,
\ee
where $A_c\equiv H_c\lambda_{\text{L}}$.

For a vacuum-superconducting interface with incident photons from the vacuum side in \emph{two} modes the current response of the superconductor in the linear response limit is to shield the EM field and reflect the radiation back into the vacuum. Thus, for photons in two modes with frequencies $\omega_1$ and $\omega_2$ the superconductor provides nearly total reflection of incident photons.
However, the nonlinear term in the current response of Eq.~\eqref{eq-Current_Nonlinear_A}, proportional to $|\vA|^2\,\vA$, leads to current sources at the vacuum-superconductor interface with frequencies, 
\be\label{eq-signal_modes}
\omega_a\in\left\{3\omega_1\,, 3\omega_2\,, 2\omega_1\pm\omega_2\,, 2\omega_2\pm\omega_1\right\}
\,,
\ee
that can radiate photons back into the vacuum. 

Consider an incident EM field with photons in modes $\omega_{1,2}$ with polarizations $\ve_{1,2}$ and amplitudes, 
\be
\vA_{1,2} = A_{1,2}(z)\,\ve_{1,2}\,\Re{e^{i\omega_{1,2}t}}
\,,
\ee
that penetrate into the London penetration depth region of the superconductor, The nonlinear term in the current response then generates current sources that can radiate photons at any of the frequencies $\omega_a$,
\be 
\vj_s(\omega_a) = \frac{c}{4\pi\lambda_{\text{L}}^2}\,\frac{\theta}{A_c^2}\vG(\omega_a)
\,,
\ee
where $\vG(\omega_a)$ are the sources generated by incident EM fields at $\omega_1$ and $\omega_2$ 
that radiate photons at the third harmonics and the intermodulation frequencies, 
\ber
\vG(3\omega_1)&=&\frac{1}{4}\,A_1^3\,\ve_1\,\Re e^{i\,3\omega_1 t}
\,,
\\
\vG(3\omega_2)&=&\frac{1}{4}\,A_2^3\,\ve_2\,\Re e^{i\,3\omega_2 t}
\,,
\qquad
\\
\vG(2\omega_1\pm\omega_2)&=&\frac{1}{2}\,A_1^2\,A_2\,\left[(\ve_1\cdot\ve_2)\ve_1+\nicefrac{1}{2}\ve_2\right]
\nonumber\\
&\times&\Re e^{i(2\omega_1\pm\omega_2) t}
\,,
\qquad
\label{eq-G_2omega1-omega2}
\\
\vG(2\omega_2\pm\omega_1)&=&\frac{1}{2}\,A_2^2\,A_1\,\left[(\ve_1\cdot\ve_2)\ve_2+\nicefrac{1}{2}\ve_1\right]
\nonumber\\
&\times&\Re e^{i(2\omega_2\pm\omega_1) t}
\,.
\eer
Note that third harmonic generation is gauranteed with a polarization state referenced to that of the relevant incident photon. For radiation at the intermodulation frequencies the polarization states of the emitted photons depends on the polarization states of both incident photons.

\subsection{Nonlinear Kerr Rotation}\label{sec-Nonlinear_Kerr_Effect}

In the linear response limit incident photons are reflected by the superconductor back into the vacuum at the same frequency, and if the superconductor is non-chiral, with the same polarization.\cite{yip92} However, if two modes with non-orthogonal ($\ve_1\cdot\ve_2\ne 0$)  and non-parallel ($\ve_1\cdot\ve_2\ne \pm 1$)  polarizations are incident on the superconductor, the nonlinear coupling generates a source current at the incident frequencies that radiates photons in a superposition of the two incident polarization states, 
\ber
\vG(\omega_1)&=&\left[\left(\frac{3}{4}\,A_1^3+\frac{1}{2}A_1\,A_2^2\right)\,\ve_1 
              + A_1\,A_2^2\,(\ve_1\cdot\ve_2)\ve_2\right]\,\Re e^{i\omega_1 t}
\,,
\qquad
\\
\vG(\omega_2)&=&\left[\left(\frac{3}{4}\,A_2^3+\frac{1}{2}A_2\,A_2^2\right)\,\ve_2 
              + A_2\,A_1^2\,(\ve_1\cdot\ve_2)\ve_1\right]\,\Re e^{i\omega_2 t}
\,.
\qquad
\eer
Thus, the nonlinear Meissner current, driven by a beam of photons in two modes, can induce a polar Kerr effect - i.e. rotation of the polarization state for the reflected photons. Observation of this effect would be a direct signature of the nonlinear Meissner current with the mixing between the two polarization states proportional to the nonlinear coupling $\theta$. 

\subsection{SRF Cavities as Axion Detectors}\label{sec-Nonlinear_SRF_Signal}

Superconducting RF cavities with state of the art quality factors,~\cite{gra13} $Q\approx 10^{11}$,
have been proposed as detectors for light weakly-coupled particles such as low-mass axions.~\cite{bog19,cao20} The idea is based on a symmetry allowed coupling of the dark-sector axion field to the visible-sector EM field, $\cL_{\text{int}}=g_{a\gamma\gamma}\,a\,\vE\cdot\vB$, where $a(\vr,t)$ is the amplitude of the axion field and $g_{a\gamma\gamma}$ is the axion-photon coupling. The combination of an axion field in the presence of a radiation field with $\vE\cdot\vB\ne 0$ can provide a current source for radiating photons with frequency $\omega_s=2\omega_1-\omega_2$ where $\omega_{1,2}$ are the frequencies of the photons in two resonant modes of the SRF cavity and $\omega_s$ is the frequency of the signal photon generated by the axion-photon coupling. If the cavity is designed such that $\omega_s$ is a resonant frequency of the cavity then there is a large mode density at the signal frequency for the detection of axion conversion.~\cite{bog19} Central to this axion detection scheme is that there are no visible-sector sources of photons at the signal frequency, $\omega_s$.
Thus, it seems essential to suppress the source current from the nonlinear screening current response at the signal frequency, e.g. $\omega_s=2\omega_1-\omega_2$. The restrictions on the cavity geometry and mode selection for the spectator photons and signal photons are that $\omega_{1,2,s}$ are all resonant modes, \emph{but} $\vG(\omega_s)\equiv 0$ for selected spectator modes of the SRF cavity geometry.\footnote{This constraint cannot be satisfied for photons at normal incidence on a plane vacuum-superconducting inferface as is evident from Eq.\eqref{eq-G_2omega1-omega2}.}
For a geometry such as that of the Tesla cavities~\cite{pad09} even a small region of the cavity surface in which $|\vG(\omega_s )|\approx A_1^2A_2$ will lead to radiation into the signal mode that will likely swamp even an optimistic estimate of the number of photons generated by axion-like dark matter if the nonlinear coupling parameter $\theta\sim \cO(0.1)$.
An analysis of the possible modes satisfying these constraints for high-Q Tesla cavities, or possibly other geometries, as well as the level of disorder that can be tolerated in order to suppress surface radiation into the signal mode, is needed.
If it is not possible select modes with $\vG(\omega_s)\equiv 0$ everywhere on the cavity surface then an alternative approach would be fabrication of an ultra-clean SRF cavity, and operation of the cavity detector at ultra-low temperatures under excitation of the two spectator modes. In the ultra-clean limit at ultra-low temperatures the nonlinear coupling is exponentially suppressed, $\theta\sim e^{-\Delta(0)/\kb T}$.~\cite{xu94a} Feasibility studies for suppressing visible-sector photons from the nonlinear surface current response at the signal frequency for axion detection are underway.

%
\section*{Acknowledgements}

This article began as a section in a review article that was to be co-authored with Dierk Rainer on {\it Quasiclassical Theory of Superconductivity in Strongly Correlated Systems}. We were not able to complete the project together before Dierk passed away. This shorter contribution is dedicated to Dierk's memory. His insights on physics were an inspiration to me, and to many other physicists who had the good fortune to know him.
The application of the strong disorder limit of quasiclassical theory to the nonlinear current response of disordered superconductors is work by the author, supported by the U.S. Department of Energy, Office of Science, National Quantum Information Science Research Centers, Superconducting Quantum Materials and Systems Center (SQMS) under contract number DE-AC02-07CH11359.
I thank my colleagues Yonatan Kahn and Roni Harnik for discussing with me their ideas for detecting axion-like dark matter using high-Q SRF cavities. 
I thank Mehdi Zarea, Hikaru Ueki and Wei-Ting Lin for discussions that improved the presentation of results reported in this manuscript.
A draft of this manuscript was written during a recent stay at the Aspen Center for Physics, which is supported by National Science Foundation grant PHY-1607611.

\appendix

\section{Propagators and Symmetries}\label{appendix-propagators}

The central feature in the theory of superconductivity is quantum mechanical coherence between normal-state particle and hole excitations. 
\emph{Particle-hole coherence} is responsible for most of the unique properties of be superconductors, including the Josephson effect, the proximity effect of superconducting-normal metal interfaces, and branch conversion scattering between particle- and hole-like excitations, i.e. \emph{Andreev scattering}.
The internal space of states for Fermions in systems with pairing correlations is the $4\times 4$ space of \emph{particle-hole $\otimes$ spin} or Nambu space.

\subsection*{Nambu Green's Functions}

Following Nambu~\cite{nam60} and Gorkov~\cite{gor59} I introduce particle-hole coherence into the Green's functions for Fermi systems by enlarging the usual two-component spinors for the creation and annihilation of spin-up and spin-down Fermions, 
\be\label{psi}
\vpsi(\vr,t)=
             \begin{pmatrix}
             \psi_{\uparrow} \cr
             \psi_{\downarrow} 
             \end{pmatrix}
\quad , \quad
\vpsi^{\dag}(\vr,t)=
             \begin{pmatrix}
		\psi^{\dag}_{\uparrow}   \cr
                \psi^{\dag}_{\downarrow}	
	     \end{pmatrix}
\,,
\ee
into four-component spinors which incorporate the particle-hole (``iso-spin'') degree of freedom,\footnote{In constructing Green's functions and propagators I form outer products of the Nambu spinors, and thus in any bilinear product such as $\Psi^{\dag}(1)\Psi(2)$ it is assumed that the spinor on the left is a column spinor while the spinor on the right is a row spinor.}
\be\label{Nambu_Psi}
\Psi(\vr,t)=\begin{pmatrix}
		\vpsi      \cr
		\vpsi^{\dag}
             \end{pmatrix}
           = \begin{pmatrix}
		\psi_{\uparrow}          \cr
                \psi_{\downarrow}        \cr
         	\psi^{\dag}_{\uparrow}   \cr
		\psi^{\dag}_{\downarrow} 
             \end{pmatrix}
\,,
\ee
\be\label{Nambu_Psi-dagger}
\Psi^{\dag}(\vr,t)=\begin{pmatrix}
			\vpsi^{\dag}       &
		        \vpsi  
		   \end{pmatrix}
                  =\begin{pmatrix}
                        \psi^{\dag}_{\uparrow}          &
                        \psi^{\dag}_{\downarrow}        &
		        \psi_{\uparrow}                 &
		        \psi_{\downarrow}
		    \end{pmatrix}
\,.
\ee
Coherent mixing of the `particle' ($\vpsi^{\dag}$) and `hole' ($\vpsi$) amplitudes is then accomplished by the off-diagonal self-energy (``pairing potential'') that fixes the amplitudes for mixing of particle and hole excitations.

It is convenient to introduce a short-hand notation for the space-time labels: $i=(\vr_i,t_i)$. Then the outer products of the four-component Nambu spinors can be written in particle-hole space as a $2\times 2$ matrix of outer products of two-component spinors,
\be
\Psi(1)\Psi^{\dag}(2) = \begin{pmatrix}
                       \vpsi(1)\vpsi^{\dag}(2)		&	\vpsi(1)\vpsi(2) \cr
                       \vpsi^{\dag}(1)\vpsi^{\dag}(2)	&	\vpsi^{\dag}(1)\vpsi(2)
	       	      \end{pmatrix}
\,,
\ee
\be
\Psi^{\dag}(2)\Psi(1) = \begin{pmatrix}
			\vpsi^{\dag}(2)\vpsi(1)		&	\vpsi(2)\vpsi(1) \cr
                        \vpsi^{\dag}(2)\vpsi^{\dag}(1)	&	\vpsi(2)\vpsi^{\dag}(1) 
		      \end{pmatrix}
\,.
\ee
The matrix structure of the four types of propagators is as follows: the retarded (R), advanced (A) and Keldysh (K) propagators encode information about the non-equilibrium excitation spectrum and occupation of states. The Matsubara (M) propagators provide information about the equilibrium quasiparticle and pairing correlations. The corresponding propagators are defined as follows,
\ber
\whG^R(1,2)
\ns&=&\ns
-i\Theta(t_1-t_2)
\left\langle\left\{\Psi(\vr_1,t_1)\,,\,\Psi^{\dag}(\vr_2,t_2)\right\}\right\rangle
\,,\quad
\\
\whG^A(1,2)
\ns&=&\ns
+i\Theta(t_2-t_1)
\left\langle\left\{\Psi(\vr_1,t_1)\,,\,\Psi^{\dag}(\vr_2,t_2)\right\}\right\rangle
\,,
\\
\whG^K(1,2)
\ns&=&\ns
-i
\left\langle\left[\Psi(\vr_1,t_1)\,,\,\Psi^{\dag}(\vr_2,t_2)\right]\right\rangle
\,,
\\
\whG^M(1,2)
\ns&=&\ns
-\left\langle\T_{\tau}\,\Psi(\vr_1,\tau_1)\,\bar{\Psi}(\vr_2,\tau_2)\right\rangle
\,,
\eer
where $[\whA,\whB] = \whA\whB-\whB\whA$ and $\{\whA,\whB\}=\whA\whB+\whB\whA$ and the expectation values are taken in a restricted Grand Canonical ensemble that allows for broken symmetry.

The Matsubara propagators depend on the imaginary-time variable, $t\rightarrow -i\tau$, confined to the strip, $-\beta\le\tau\le\beta$. Note that the adjoint is defined for imaginary times by $\bar{\Psi}(\vr,\tau)\equiv\Psi^{\dag}(-\tau)$. Finally, the imaginary-time ordering operation is defined as
\be
\T_{\tau}\left(\Psi(1)\,\bar{\Psi}(2)\right) = 
\Bigg\{
\begin{matrix}
	+\Psi(1)\,\bar{\Psi}(2)\,,\quad\tau_1 > \tau_2\, \cr
        -\bar{\Psi}(2)\,\Psi(1)\,,\quad\tau_1 < \tau_2\,.
\end{matrix}
\ee

The compact Nambu matrix notation can be expanded as a $2\times 2$ matrix in particle-hole space,
\be
\whG^{x}(1,2) = \begin{pmatrix} 
			G^x(1,2)	& 	F^x(1,2) \cr
		   \bar{F}^x(1,2)	& 	\bar{G}^x(1,2)
		   \end{pmatrix}
\,,
\ee
with $x=(R,A,K,M)$ and each entry being a $2\times 2$ spin matrix. In particular, $G^x$ and $\bar{G}^x$ are the `conventional' diagonal Green's functions and $F^x$ and $\bar{F}^x$ are the anomalous (Gorkov) functions which define the pairing correlations of the superconducting state. 

\subsection*{Mixed Representation}

The most convenient set of variables in which to express the propagators is the mixed-representation obtained by first transforming to center-of-mass and relative space-time coordinates,
\ber
\vR=(\vr_1+\vr_2)/2\,&,&\quad\vr=\vr_1-\vr_2
\\
t=(t_1+t_2)/2\,&,&\quad s=t_1-t_2
\,,
\eer
then Fourier transforming with respect to the relative coordinates,
\ber
\whG^{x}(\vp,\varepsilon;\vR,t)
=
\int\,d^3r\,\int_{-\infty}^{+\infty}ds\,e^{-i(\vp\cdot\vr -\varepsilon s)}\,
\whG^{x}(\vR+\vr/2,t+s/2;\vR-\vr/2,t-s/2)
\,,
\eer
for $x=(R,A,K)$. Similary, for the Matsubara propagator in the mixed representation I obtain,
\ber
\whG^{M}(\vp,\varepsilon_n;\vR)
=
\int\,d^3r\,\int_{-\beta}^{+\beta}d\tau\,e^{-i(\vp\cdot\vr -\varepsilon_n\tau)}\,
\whG^{M}(\vR+\vr/2,\tau;\vR-\vr/2,0) 
\,,
\eer
where $\varepsilon_n=(2n+1)\pi k_B T$ are the Fermion Matsubara frequencies.

\subsection*{Quasiclassical Propagators}

Finally I define the Nambu matrix form for the quasiclassical propagators, obtained by integrating the full Green's functions with respect to the magnitude of the momentum perpendicular to the Fermi surface, or equivalently over the normal-state excitation energy,
\be
\xi_{\vp}=v_{\vp}\left(\vert\vp\vert - p_f\right)
\,.
\ee
The quasiclassical propagators depend on the position \emph{on the Fermi surface} defined by the Fermi momentum, $\vp$. It is also conventional to pre-multiply the full propagators by the Nambu matrix
\be
\tz=\begin{pmatrix}\vone & 0 \cr 0 & -\vone\end{pmatrix}
\,,
\ee
in particle-hole space,\footnote{This definition is connected with Eilenberger's reduction of Gorkov's equations to transport-type equations.} and to renormalize the propagators by dividing by the spectral weight of the normal-state quasiparticle pole. Thus,
\ber\label{eq-quasiclassical_propagators}
\whmfG^{x}(\vp,\varepsilon;\vR,t)
&\equiv&
\frac{1}{a}\int\,d\xi_{\vp}\,\tz\,\whG^x(\vp,\varepsilon;\vR,t)
\,,\qquad
\\
\therefore\,
\whmfG^{x}
=
\begin{pmatrix} \mfg^{x} & \mff^{x} \cr \mfuf^{x} & \mfug^{x} \end{pmatrix}
&=&
\frac{1}{a}\int\,d\xi_{\vp}\,
\begin{pmatrix} G^{x} & F^{x} \cr -\bar{F}^{x} & -\bar{G}^{x} \end{pmatrix}
\,,\qquad
\eer
are the quasiclassical propagators that satisfy the Eilenberger transport equations and normalization conditions. Note that the momenta are evaluated at the Fermi surface on the left side of Eq.~\ref{eq-quasiclassical_propagators} for all quasiclassical propgators.

\subsection*{Particle-Hole and Conjugation Symmetries}

By using fundamental symmetries under particle-hole (conjugation) and Fermion exchange I can derive the following matrix identities relating components of the quasiclassical propagators,

\ber
\whmfGra(\vp,\varepsilon;\vR,t) &=& +\Big[\tx\whmfGra(-\vp,-\varepsilon;\vR,t)^{*}\tx\Big]
\,,\qquad
\label{gRA_conjugation}
\\
\whmfGk(\vp,\varepsilon;\vR,t) &=&-\Big[\tx\whmfGk(-\vp,-\varepsilon;\vR,t)^{*}\tx\Big]
\,,
\label{gK_conjugation}
\\
\whmfGm(\vp,\varepsilon_n;\vR) &=&+\Big[\tx\whmfGm(-\vp,+\varepsilon_n;\vR)^{*}\tx\Big]
\,,
\label{gM_conjugation}
\eer

\ber
\whmfGr(\vp,\varepsilon;\vR,t) &=&+\Big[\tz\whmfGa(\vp,\varepsilon;\vR,t)^{\dag}\tz\Big]
\,,\qquad
\label{gRA_combined}
\\
\whmfGk(\vp,\varepsilon;\vR,t) &=&-\Big[\tz\whmfGk(\vp,\varepsilon;\vR,t)^{\dag}\tz\Big]
\,,
\label{gK_combined}
\\
\whmfGm(\vp,\varepsilon_n;\vR) &=&+\Big[\tz\whmfGm(\vp,-\varepsilon_n;\vR)^{\dag}\tz\Big]
\,.
\label{gM_combined}
\eer

\section{Nambu Matrix Algebra}\label{appendix-Nambu_Algebra}

The Nambu matrices obey the algebra of the generators of $\point{SU(2)}{\ns}$,
\be
\widehat{\tau}_i\,\widehat{\tau}_j=\delta_{ij}\,\tone + i\,\epsilon_{ijk}\,
                  \widehat{\tau}_{k}\,,\quad i,j\in\{x,y,z\}
\,.
\ee
In the ``circular'' basis,
\ber
\tp\,\tm = \nicefrac{1}{2}\left(\tone + \tz\right)
\,,
&\quad&
\tm\,\tp = \nicefrac{1}{2}\left(\tone - \tz\right)
\,,
\\
\commutator{\tp}{\tm} = \tz
\,,
\quad
&\quad&
\anticommutator{\tp}{\tm} = \tone
\,,
\\
\commutator{\tz}{\tpm} = \pm 2\,\tpm
\,,
&\quad&
\anticommutator{\tz}{\tpm} = 0
\,,
\eer
where $\commutator{\whmfa}{\whmfb}=\whmfa\whmfb - \whmfb\whmfa$ and 
$\anticommutator{\whmfa}{\whmfb}=\whmfa\whmfb + \whmfb\whmfa$.

\begin{figure}[t]
\centering\includegraphics[width=0.65\columnwidth]{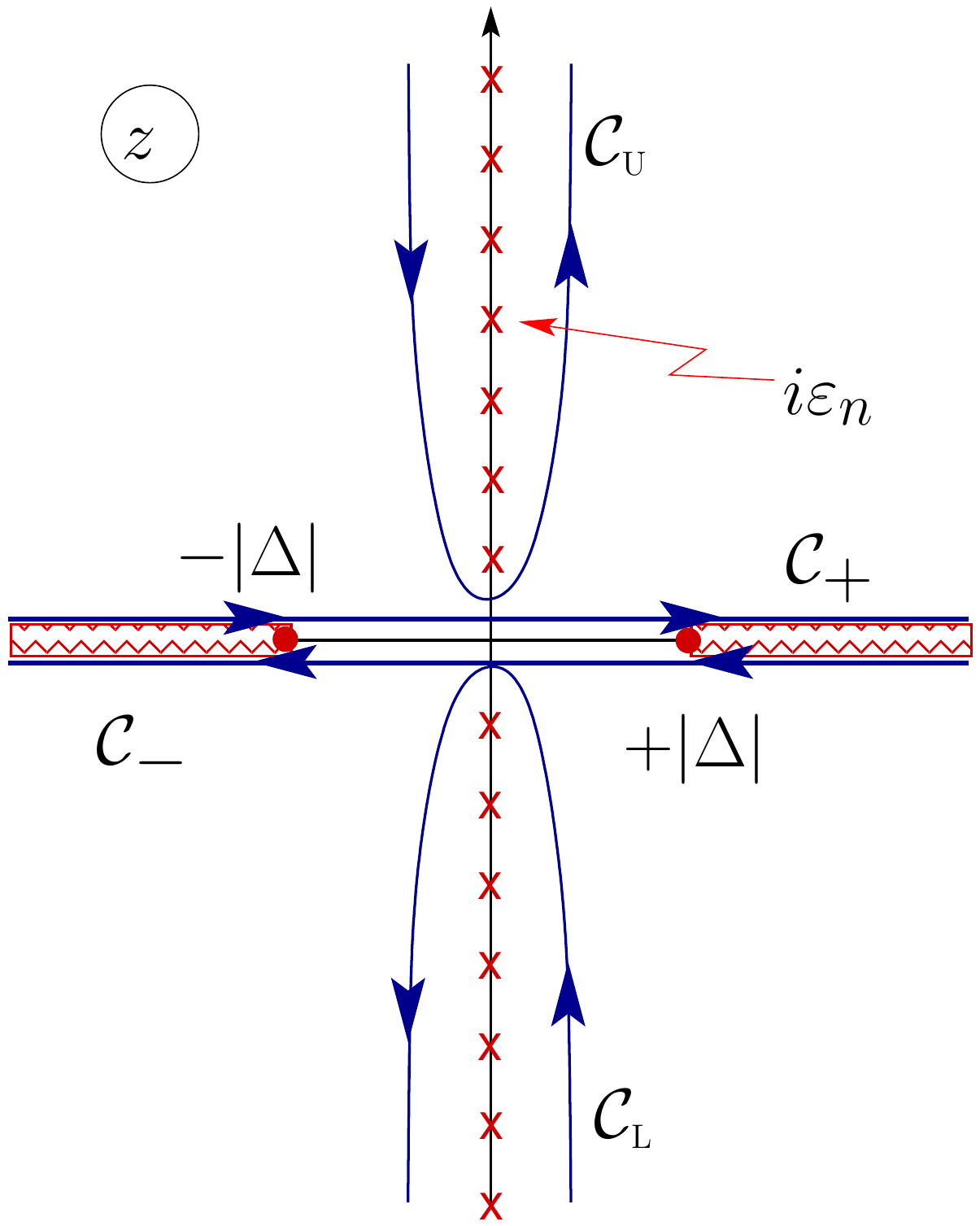}
\caption{
Contour integration for analytic continuation to real energies. In the dirty limit the spectral current density is dominated by poles at $\varepsilon_{\pm}=\pm|\Delta|$.}\label{fig-Matsubara_Transformation}
\end{figure}

\section{Analytic Continuation}\label{appendix-Matsubara_Transformation}

Matsubara sums for equilibrium properties such as the current in Eq. \ref{eq-Current_Matsubara} can be transformed to an integration over all energies weighted by an appropriate spectral density and the Fermi distribution function. In particular, consider transforming the sum $T\sum_n G(\varepsilon_n)$ to an integration over the real axis. First, Cauchy's theorem allows us to represent the Matsubara sum as a contour integral enclosing all the poles of the Fermi function as shown in Fig. \ref{fig-Matsubara_Transformation},
\ber
T\sum_{\varepsilon_n}\,G(\varepsilon_n) 
&=& 
-\frac{1}{2\pi i} \oint_{\cC_{\text{L}}+\cC_{\text{U}}}\,dz\,f(z)\,G(z)
\,,
\eer
provided $G(z)$ is analytic in the neighborhood of the imaginary axis. Then, if $G(z)$ is piecewise analytic function in the upper and lower half plane, \emph{and} vanishes faster than $1/z$ for $|z|\rightarrow\infty$, I transform the integration over contours $\cC_{\text{L}}+\cC_{\text{U}}$ to the contours $\cC_{\pm}$ just above and below the real axis, and thus to a single integration along the real axis,
\ber
T\sum_{\varepsilon_n}\ns G(\varepsilon_n) 
\ns&=&\ns 
-\frac{1}{2\pi i}\int_{-\infty}^{+\infty}\ns\ns d\varepsilon\,f(\varepsilon)\,
                 \left[G^{\text{R}}(\varepsilon) - G^{\text{A}}(\varepsilon)\right],
\quad
\eer
where $G^{\text{R/A}}(\varepsilon) = G(z\rightarrow\varepsilon \pm i 0^+)$.


\begin{thebibliography}{41}%
\makeatletter
\providecommand \@ifxundefined [1]{%
 \@ifx{#1\undefined}
}%
\providecommand \@ifnum [1]{%
 \ifnum #1\expandafter \@firstoftwo
 \else \expandafter \@secondoftwo
 \fi
}%
\providecommand \@ifx [1]{%
 \ifx #1\expandafter \@firstoftwo
 \else \expandafter \@secondoftwo
 \fi
}%
\providecommand \natexlab [1]{#1}%
\providecommand \enquote  [1]{``#1''}%
\providecommand \bibnamefont  [1]{#1}%
\providecommand \bibfnamefont [1]{#1}%
\providecommand \citenamefont [1]{#1}%
\providecommand \href@noop [0]{\@secondoftwo}%
\providecommand \href [0]{\begingroup \@sanitize@url \@href}%
\providecommand \@href[1]{\@@startlink{#1}\@@href}%
\providecommand \@@href[1]{\endgroup#1\@@endlink}%
\providecommand \@sanitize@url [0]{\catcode `\\12\catcode `\$12\catcode
  `\&12\catcode `\#12\catcode `\^12\catcode `\_12\catcode `\%12\relax}%
\providecommand \@@startlink[1]{}%
\providecommand \@@endlink[0]{}%
\providecommand \url  [0]{\begingroup\@sanitize@url \@url }%
\providecommand \@url [1]{\endgroup\@href {#1}{\urlprefix }}%
\providecommand \urlprefix  [0]{URL }%
\providecommand \Eprint [0]{\href }%
\providecommand \doibase [0]{http://dx.doi.org/}%
\providecommand \selectlanguage [0]{\@gobble}%
\providecommand \bibinfo  [0]{\@secondoftwo}%
\providecommand \bibfield  [0]{\@secondoftwo}%
\providecommand \translation [1]{[#1]}%
\providecommand \BibitemOpen [0]{}%
\providecommand \bibitemStop [0]{}%
\providecommand \bibitemNoStop [0]{.\EOS\space}%
\providecommand \EOS [0]{\spacefactor3000\relax}%
\providecommand \BibitemShut  [1]{\csname bibitem#1\endcsname}%
\let\auto@bib@innerbib\@empty
\bibitem [{\citenamefont {Landau}(1956)}]{lan56}%
  \BibitemOpen
  \bibfield  {author} {\bibinfo {author} {\bibfnamefont {L.~D.}\ \bibnamefont
  {Landau}},\ }\bibfield  {{The Theory of a Fermi Liquid}} {\emph {\bibinfo
  {title} {{The Theory of a Fermi Liquid}},\ }}\href@noop {} {\bibfield
  {journal} {\bibinfo  {journal} {Sov. Phys. JETP}\ }\textbf {\bibinfo {volume}
  {30}},\ \bibinfo {pages} {1058} (\bibinfo {year} {1956})}\BibitemShut
  {NoStop}%
\bibitem [{\citenamefont {Landau}(1957)}]{lan57}%
  \BibitemOpen
  \bibfield  {author} {\bibinfo {author} {\bibfnamefont {L.~D.}\ \bibnamefont
  {Landau}},\ }\bibfield  {{Oscillations in a Fermi Liquid}} {\emph {\bibinfo
  {title} {{Oscillations in a Fermi Liquid}},\ }}\href@noop {} {\bibfield
  {journal} {\bibinfo  {journal} {Sov. Phys. JETP}\ }\textbf {\bibinfo {volume}
  {32}},\ \bibinfo {pages} {59} (\bibinfo {year} {1957})}\BibitemShut {NoStop}%
\bibitem [{\citenamefont {Eliashberg}(1962)}]{eli62}%
  \BibitemOpen
  \bibfield  {author} {\bibinfo {author} {\bibfnamefont {G.~M.}\ \bibnamefont
  {Eliashberg}},\ }\bibfield  {{Transport Equation for a Degenerate System of
  Fermi Particles}} {\emph {\bibinfo {title} {{Transport Equation for a
  Degenerate System of Fermi Particles}},\ }}\href@noop {} {\bibfield
  {journal} {\bibinfo  {journal} {Sov. Phys. JETP}\ }\textbf {\bibinfo {volume}
  {14}},\ \bibinfo {pages} {886} (\bibinfo {year} {1962})},\ \bibinfo {note}
  {[ZhETF, 41, 1241, (1962)]}\BibitemShut {NoStop}%
\bibitem [{\citenamefont {Prange}\ and\ \citenamefont
  {Kadanoff}(1964)}]{pra64}%
  \BibitemOpen
  \bibfield  {author} {\bibinfo {author} {\bibfnamefont {R.~E.}\ \bibnamefont
  {Prange}}\ and\ \bibinfo {author} {\bibfnamefont {L.~P.}\ \bibnamefont
  {Kadanoff}},\ }\bibfield  {{Transport Theory for Electron-Phonon Interactions
  in Metals}} {\emph {\bibinfo {title} {{Transport Theory for Electron-Phonon
  Interactions in Metals}},\ }}\href {\doibase 10.1103/PhysRev.134.A566}
  {\bibfield  {journal} {\bibinfo  {journal} {Phys. Rev.}\ }\textbf {\bibinfo
  {volume} {134}},\ \bibinfo {pages} {A566} (\bibinfo {year}
  {1964})}\BibitemShut {NoStop}%
\bibitem [{\citenamefont {Baym}\ and\ \citenamefont {Pethick}(1991)}]{baym91}%
  \BibitemOpen
  \bibfield  {author} {\bibinfo {author} {\bibfnamefont {G.}~\bibnamefont
  {Baym}}\ and\ \bibinfo {author} {\bibfnamefont {C.~J.}\ \bibnamefont
  {Pethick}},\ }\href@noop {} {\emph {\bibinfo {title} {{Landau Fermi-Liquid
  Theory}}}}\ (\bibinfo  {publisher} {Wiley},\ \bibinfo {address} {New York},\
  \bibinfo {year} {1991})\BibitemShut {NoStop}%
\bibitem [{\citenamefont {Bardeen}\ \emph {et~al.}(1957)\citenamefont
  {Bardeen}, \citenamefont {Cooper},\ and\ \citenamefont {Schrieffer}}]{bar57}%
  \BibitemOpen
  \bibfield  {author} {\bibinfo {author} {\bibfnamefont {J.}~\bibnamefont
  {Bardeen}}, \bibinfo {author} {\bibfnamefont {L.}~\bibnamefont {Cooper}}, \
  and\ \bibinfo {author} {\bibfnamefont {J.}~\bibnamefont {Schrieffer}},\
  }\bibfield  {{Theory of Superconductivity}} {\emph {\bibinfo {title} {{Theory
  of Superconductivity}},\ }}\href {\doibase 10.1103/PhysRev.108.1175}
  {\bibfield  {journal} {\bibinfo  {journal} {Phys. Rev.}\ }\textbf {\bibinfo
  {volume} {108}},\ \bibinfo {pages} {1175} (\bibinfo {year}
  {1957})}\BibitemShut {NoStop}%
\bibitem [{\citenamefont {Bogoliubov}(1958)}]{bog58a}%
  \BibitemOpen
  \bibfield  {author} {\bibinfo {author} {\bibfnamefont {N.~N.}\ \bibnamefont
  {Bogoliubov}},\ }\bibfield  {{New Method in the Theory of Superconductivity}}
  {\emph {\bibinfo {title} {{New Method in the Theory of Superconductivity}},\
  }}\href@noop {} {\bibfield  {journal} {\bibinfo  {journal} {Zh. Eskp. Teor.
  Fiz.}\ }\textbf {\bibinfo {volume} {34}},\ \bibinfo {pages} {58} (\bibinfo
  {year} {1958})},\ \bibinfo {note} {[english: Sov. Phys. JETP 7, 41-46
  (1958)]}\BibitemShut {NoStop}%
\bibitem [{\citenamefont {Gorkov}(1959)}]{gor59}%
  \BibitemOpen
  \bibfield  {author} {\bibinfo {author} {\bibfnamefont {L.}~\bibnamefont
  {Gorkov}},\ }\bibfield  {{Microscopic Derivation of the Ginzburg-Landau
  Equations in the Theory of Superconductivity}} {\emph {\bibinfo {title}
  {{Microscopic Derivation of the Ginzburg-Landau Equations in the Theory of
  Superconductivity}},\ }}\href@noop {} {\bibfield  {journal} {\bibinfo
  {journal} {Sov. Phys. JETP}\ }\textbf {\bibinfo {volume} {9}},\ \bibinfo
  {pages} {1364} (\bibinfo {year} {1959})}\BibitemShut {NoStop}%
\bibitem [{\citenamefont {Eilenberger}(1968)}]{eil68}%
  \BibitemOpen
  \bibfield  {author} {\bibinfo {author} {\bibfnamefont {G.}~\bibnamefont
  {Eilenberger}},\ }\bibfield  {{Transformation of Gorkov's Equation for Type
  II Superconductors into Transport-Like Equations}} {\emph {\bibinfo {title}
  {{Transformation of Gorkov's Equation for Type II Superconductors into
  Transport-Like Equations}},\ }}\href {\doibase 10.1007/BF01379803} {\bibfield
   {journal} {\bibinfo  {journal} {Zeit. f. Physik}\ }\textbf {\bibinfo
  {volume} {214}},\ \bibinfo {pages} {195} (\bibinfo {year}
  {1968})}\BibitemShut {NoStop}%
\bibitem [{\citenamefont {Larkin}\ and\ \citenamefont
  {Ovchinnikov}(1969)}]{lar69}%
  \BibitemOpen
  \bibfield  {author} {\bibinfo {author} {\bibfnamefont {A.~I.}\ \bibnamefont
  {Larkin}}\ and\ \bibinfo {author} {\bibfnamefont {Y.~N.}\ \bibnamefont
  {Ovchinnikov}},\ }\bibfield  {{Quasiclassical Method in the Theory of
  Superconductivity}} {\emph {\bibinfo {title} {{Quasiclassical Method in the
  Theory of Superconductivity}},\ }}\href@noop {} {\bibfield  {journal}
  {\bibinfo  {journal} {Sov. Phys. JETP}\ }\textbf {\bibinfo {volume} {28}},\
  \bibinfo {pages} {1200} (\bibinfo {year} {1969})}\BibitemShut {NoStop}%
\bibitem [{\citenamefont {Eliashberg}(1972)}]{eli72}%
  \BibitemOpen
  \bibfield  {author} {\bibinfo {author} {\bibfnamefont {G.~M.}\ \bibnamefont
  {Eliashberg}},\ }\bibfield  {{Inelastic electron collisions and
  nonequilibrium stationary states in superconductors}} {\emph {\bibinfo
  {title} {{Inelastic electron collisions and nonequilibrium stationary states
  in superconductors}},\ }}\href@noop {} {\bibfield  {journal} {\bibinfo
  {journal} {Sov. Phys. JETP}\ }\textbf {\bibinfo {volume} {34}},\ \bibinfo
  {pages} {668} (\bibinfo {year} {1972})}\BibitemShut {NoStop}%
\bibitem [{\citenamefont {Larkin}\ and\ \citenamefont
  {Ovchinnikov}(1975)}]{lar75}%
  \BibitemOpen
  \bibfield  {author} {\bibinfo {author} {\bibfnamefont {A.}~\bibnamefont
  {Larkin}}\ and\ \bibinfo {author} {\bibfnamefont {Y.}~\bibnamefont
  {Ovchinnikov}},\ }\bibfield  {{Nonlinear conductivity of superconductors in
  the mixed state}} {\emph {\bibinfo {title} {{Nonlinear conductivity of
  superconductors in the mixed state}},\ }}\href@noop {} {\bibfield  {journal}
  {\bibinfo  {journal} {Sov. Phys. JETP}\ }\textbf {\bibinfo {volume} {41}},\
  \bibinfo {pages} {960} (\bibinfo {year} {1975})}\BibitemShut {NoStop}%
\bibitem [{\citenamefont {Rainer}\ and\ \citenamefont {Sauls}(2018)}]{rai94}%
  \BibitemOpen
  \bibfield  {author} {\bibinfo {author} {\bibfnamefont {D.}~\bibnamefont
  {Rainer}}\ and\ \bibinfo {author} {\bibfnamefont {J.~A.}\ \bibnamefont
  {Sauls}},\ }\bibfield  {{Strong-Coupling Theory of Superconductivity}} {\emph
  {\bibinfo {title} {{Strong-Coupling Theory of Superconductivity}},\
  }}\bibfield  {booktitle} {\emph {\bibinfo {booktitle} {Superconductivity:
  From Basic Physics to New Developments}},\ }\href
  {https://arxiv.org/abs/1809.05264} {\bibfield  {journal} {\bibinfo  {journal}
  {arXiv:}\ }\textbf {\bibinfo {volume} {1809.05264}},\ \bibinfo {pages} {1}
  (\bibinfo {year} {2018})},\ \bibinfo {note} {published in
  ``Superconductivity: From Basic Physics to New Developments'', ch. 2, pp.
  45-78, World Scientific, Singapore (1994).}\BibitemShut {Stop}%
\bibitem [{\citenamefont {Usadel}(1970)}]{usa70}%
  \BibitemOpen
  \bibfield  {author} {\bibinfo {author} {\bibfnamefont {K.}~\bibnamefont
  {Usadel}},\ }\bibfield  {Generalized Diffusion Equation for Superconducting
  Alloys} {\emph {\bibinfo {title} {Generalized diffusion equation for
  superconducting alloys},\ }}\href {\doibase 10.1103/PhysRevLett.25.507}
  {\bibfield  {journal} {\bibinfo  {journal} {Phys. Rev. Lett.}\ }\textbf
  {\bibinfo {volume} {25}},\ \bibinfo {pages} {507} (\bibinfo {year}
  {1970})}\BibitemShut {NoStop}%
\bibitem [{\citenamefont {Kamenev}(2011)}]{kamenev11}%
  \BibitemOpen
  \bibfield  {author} {\bibinfo {author} {\bibfnamefont {A.}~\bibnamefont
  {Kamenev}},\ }\href@noop {} {\emph {\bibinfo {title} {{Field Theory of
  Non-Equilibrium Systems}}}}\ (\bibinfo  {publisher} {Cambridge University
  Press},\ \bibinfo {address} {Cambridge, UK},\ \bibinfo {year}
  {2011})\BibitemShut {NoStop}%
\bibitem [{\citenamefont {Rammer}\ and\ \citenamefont {Smith}(1986)}]{ram86}%
  \BibitemOpen
  \bibfield  {author} {\bibinfo {author} {\bibfnamefont {J.}~\bibnamefont
  {Rammer}}\ and\ \bibinfo {author} {\bibfnamefont {H.}~\bibnamefont {Smith}},\
  }\bibfield  {{Quantum field-theoretical methods in transport theory of
  metals}} {\emph {\bibinfo {title} {{Quantum field-theoretical methods in
  transport theory of metals}},\ }}\href {\doibase 10.1103/RevModPhys.58.323}
  {\bibfield  {journal} {\bibinfo  {journal} {Rev. Mod. Phys.}\ }\textbf
  {\bibinfo {volume} {58}},\ \bibinfo {pages} {323} (\bibinfo {year}
  {1986})}\BibitemShut {NoStop}%
\bibitem [{\citenamefont {Sauls}(2018)}]{sau18}%
  \BibitemOpen
  \bibfield  {author} {\bibinfo {author} {\bibfnamefont {J.~A.}\ \bibnamefont
  {Sauls}},\ }\bibfield  {{Andreev Bound States and Their Signatures}} {\emph
  {\bibinfo {title} {{Andreev Bound States and Their Signatures}},\ }}\href
  {\doibase 10.1098/rsta.2018.0140} {\bibfield  {journal} {\bibinfo  {journal}
  {Phil. Trans. Roy. Soc. A}\ }\textbf {\bibinfo {volume} {376}} (\bibinfo
  {year} {2018}),\ 10.1098/rsta.2018.0140}\BibitemShut {NoStop}%
\bibitem [{\citenamefont {Serene}\ and\ \citenamefont {Rainer}(1983)}]{ser83}%
  \BibitemOpen
  \bibfield  {author} {\bibinfo {author} {\bibfnamefont {J.~W.}\ \bibnamefont
  {Serene}}\ and\ \bibinfo {author} {\bibfnamefont {D.}~\bibnamefont
  {Rainer}},\ }\bibfield  {{The Quasiclassical Approach to $^3He$}} {\emph
  {\bibinfo {title} {{The Quasiclassical Approach to $^3He$}},\ }}\href
  {\doibase 10.1016/0370-1573(83)90051-0} {\bibfield  {journal} {\bibinfo
  {journal} {Phys. Rep.}\ }\textbf {\bibinfo {volume} {101}},\ \bibinfo {pages}
  {221} (\bibinfo {year} {1983})}\BibitemShut {NoStop}%
\bibitem [{\citenamefont {Rainer}(1986)}]{rai86}%
  \BibitemOpen
  \bibfield  {author} {\bibinfo {author} {\bibfnamefont {D.}~\bibnamefont
  {Rainer}},\ }\bibinfo {title} {{Principles of {\it ab} Initio Calculations of
  Superconducting Transition Temperatures}},\ in\ \href {\doibase
  10.1016/S0079-6417(08)60024-4} {\emph {\bibinfo {booktitle} {Progress in Low
  Temperature Physics}}},\ Vol.~\bibinfo {volume} {10}\ (\bibinfo  {publisher}
  {Elsevier Science Publishers B.V.},\ \bibinfo {address} {Amsterdam},\
  \bibinfo {year} {1986})\ pp.\ \bibinfo {pages} {371--424}\BibitemShut
  {NoStop}%
\bibitem [{\citenamefont {Smith}\ and\ \citenamefont
  {{H{\o}jgaard-Jensen}}(1989)}]{smith89}%
  \BibitemOpen
  \bibfield  {author} {\bibinfo {author} {\bibfnamefont {H.}~\bibnamefont
  {Smith}}\ and\ \bibinfo {author} {\bibfnamefont {H.}~\bibnamefont
  {{H{\o}jgaard-Jensen}}},\ }\href@noop {} {\emph {\bibinfo {title} {Transport
  Phenomena}}}\ (\bibinfo  {publisher} {Clarendon Press},\ \bibinfo {address}
  {Oxford},\ \bibinfo {year} {1989})\BibitemShut {NoStop}%
\bibitem [{\citenamefont {Schmid}\ and\ \citenamefont {Sch\"on}(1975)}]{sch75}%
  \BibitemOpen
  \bibfield  {author} {\bibinfo {author} {\bibfnamefont {A.}~\bibnamefont
  {Schmid}}\ and\ \bibinfo {author} {\bibfnamefont {G.}~\bibnamefont
  {Sch\"on}},\ }\bibfield  {{Collective Oscillations in a Dirty
  Superconductor}} {\emph {\bibinfo {title} {{Collective Oscillations in a
  Dirty Superconductor}},\ }}\href {\doibase 10.1103/PhysRevLett.34.941}
  {\bibfield  {journal} {\bibinfo  {journal} {Phys. Rev. Lett.}\ }\textbf
  {\bibinfo {volume} {34}},\ \bibinfo {pages} {941} (\bibinfo {year}
  {1975})}\BibitemShut {NoStop}%
\bibitem [{\citenamefont {Alexander}\ \emph {et~al.}(1985)\citenamefont
  {Alexander}, \citenamefont {Orlando}, \citenamefont {Rainer},\ and\
  \citenamefont {Tedrow}}]{ale85}%
  \BibitemOpen
  \bibfield  {author} {\bibinfo {author} {\bibfnamefont {J.}~\bibnamefont
  {Alexander}}, \bibinfo {author} {\bibfnamefont {T.}~\bibnamefont {Orlando}},
  \bibinfo {author} {\bibfnamefont {D.}~\bibnamefont {Rainer}}, \ and\ \bibinfo
  {author} {\bibfnamefont {P.}~\bibnamefont {Tedrow}},\ }\bibfield  {{Theory of
  Fermi-liquid effects in high-field tunneling}} {\emph {\bibinfo {title}
  {{Theory of Fermi-liquid effects in high-field tunneling}},\ }}\href
  {\doibase 10.1103/PhysRevB.31.5811} {\bibfield  {journal} {\bibinfo
  {journal} {Phys. Rev. B}\ }\textbf {\bibinfo {volume} {31}},\ \bibinfo
  {pages} {5811} (\bibinfo {year} {1985})}\BibitemShut {NoStop}%
\bibitem [{\citenamefont {Anderson}(1959)}]{and59}%
  \BibitemOpen
  \bibfield  {author} {\bibinfo {author} {\bibfnamefont {P.~W.}\ \bibnamefont
  {Anderson}},\ }\bibfield  {{Theory of Dirty Superconductors}} {\emph
  {\bibinfo {title} {{Theory of Dirty Superconductors}},\ }}\href {\doibase
  10.1016/0022-3697(59)90036-8} {\bibfield  {journal} {\bibinfo  {journal} {J.
  Phys. Chem. Sol.}\ }\textbf {\bibinfo {volume} {11}},\ \bibinfo {pages} {26 }
  (\bibinfo {year} {1959})}\BibitemShut {NoStop}%
\bibitem [{\citenamefont {Shelankov}(1980)}]{she80}%
  \BibitemOpen
  \bibfield  {author} {\bibinfo {author} {\bibfnamefont {A.}~\bibnamefont
  {Shelankov}},\ }\bibfield  {Dragging of normal component by the condensate in
  nonequimbrium superconductors} {\emph {\bibinfo {title} {Dragging of normal
  component by the condensate in nonequimbrium superconductors},\ }}\href@noop
  {} {\bibfield  {journal} {\bibinfo  {journal} {Sov. Phys. JETP}\ }\textbf
  {\bibinfo {volume} {51}},\ \bibinfo {pages} {1186} (\bibinfo {year}
  {1980})}\BibitemShut {NoStop}%
\bibitem [{\citenamefont {Fomin}(2018)}]{fom18}%
  \BibitemOpen
  \bibfield  {author} {\bibinfo {author} {\bibfnamefont {I.~A.}\ \bibnamefont
  {Fomin}},\ }\bibfield  {{Analog of the Anderson Theorem for the Polar Phase
  of Liquid $^3$He in a Nematic Aerogel}} {\emph {\bibinfo {title} {{Analog of
  the Anderson Theorem for the Polar Phase of Liquid $^3$He in a Nematic
  Aerogel}},\ }}\href {\doibase 10.1134/S106377611811002X} {\bibfield
  {journal} {\bibinfo  {journal} {Sov. Phys. JETP}\ }\textbf {\bibinfo {volume}
  {127}},\ \bibinfo {pages} {933} (\bibinfo {year} {2018})}\BibitemShut
  {NoStop}%
\bibitem [{\citenamefont {Abrikosov}\ and\ \citenamefont
  {Gorkov}(1959{\natexlab{a}})}]{abr59b}%
  \BibitemOpen
  \bibfield  {author} {\bibinfo {author} {\bibfnamefont {A.~A.}\ \bibnamefont
  {Abrikosov}}\ and\ \bibinfo {author} {\bibfnamefont {L.~P.}\ \bibnamefont
  {Gorkov}},\ }\bibfield  {{Superconducting Alloys at Finite Temperatures}}
  {\emph {\bibinfo {title} {{Superconducting Alloys at Finite Temperatures}},\
  }}\href@noop {} {\bibfield  {journal} {\bibinfo  {journal} {Sov. Phys. JETP}\
  }\textbf {\bibinfo {volume} {9}},\ \bibinfo {pages} {220} (\bibinfo {year}
  {1959}{\natexlab{a}})}\BibitemShut {NoStop}%
\bibitem [{\citenamefont {Edwards}(1958)}]{edw58}%
  \BibitemOpen
  \bibfield  {author} {\bibinfo {author} {\bibfnamefont {S.~F.}\ \bibnamefont
  {Edwards}},\ }\bibfield  {{A New Method for the Evaluation of the Electrical
  Conductivity in Metals}} {\emph {\bibinfo {title} {{A New Method for the
  Evaluation of the Electrical Conductivity in Metals}},\ }}\href {\doibase
  10.1080/14786435808243244} {\bibfield  {journal} {\bibinfo  {journal} {Phil.
  Mag.}\ }\textbf {\bibinfo {volume} {3}},\ \bibinfo {pages} {1020} (\bibinfo
  {year} {1958})}\BibitemShut {NoStop}%
\bibitem [{\citenamefont {Markowitz}\ and\ \citenamefont
  {Kadanoff}(1963)}]{mar63}%
  \BibitemOpen
  \bibfield  {author} {\bibinfo {author} {\bibfnamefont {D.}~\bibnamefont
  {Markowitz}}\ and\ \bibinfo {author} {\bibfnamefont {L.~P.}\ \bibnamefont
  {Kadanoff}},\ }\bibfield  {{Effect of Impurities upon Critical Temperature of
  Anisotropic Superconductors}} {\emph {\bibinfo {title} {{Effect of Impurities
  upon Critical Temperature of Anisotropic Superconductors}},\ }}\href
  {\doibase 10.1103/PhysRev.131.563} {\bibfield  {journal} {\bibinfo  {journal}
  {Phys. Rev.}\ }\textbf {\bibinfo {volume} {131}},\ \bibinfo {pages} {563}
  (\bibinfo {year} {1963})}\BibitemShut {NoStop}%
\bibitem [{\citenamefont {Hohenberg}(1964)}]{hoh64}%
  \BibitemOpen
  \bibfield  {author} {\bibinfo {author} {\bibfnamefont {P.}~\bibnamefont
  {Hohenberg}},\ }\bibfield  {Anisotropic Superconductors with NonMagnetic
  Impurities} {\emph {\bibinfo {title} {Anisotropic superconductors with
  nonmagnetic impurities},\ }}\href@noop {} {\bibfield  {journal} {\bibinfo
  {journal} {Sov. Phys. JETP}\ }\textbf {\bibinfo {volume} {18}},\ \bibinfo
  {pages} {834} (\bibinfo {year} {1964})}\BibitemShut {NoStop}%
\bibitem [{\citenamefont {Zarea}\ \emph {et~al.}(2022)\citenamefont {Zarea},
  \citenamefont {Ueki},\ and\ \citenamefont {Sauls}}]{zar22}%
  \BibitemOpen
  \bibfield  {author} {\bibinfo {author} {\bibfnamefont {M.}~\bibnamefont
  {Zarea}}, \bibinfo {author} {\bibfnamefont {H.}~\bibnamefont {Ueki}}, \ and\
  \bibinfo {author} {\bibfnamefont {J.~A.}\ \bibnamefont {Sauls}},\ }\bibfield
  {{Effects of anisotropy and disorder on the superconducting properties of
  Niobium}} {\emph {\bibinfo {title} {{Effects of anisotropy and disorder on
  the superconducting properties of Niobium}},\ }}\href
  {https://arxiv.org/abs/2201.07403} {\bibfield  {journal} {\bibinfo  {journal}
  {arXiv}\ }\textbf {\bibinfo {volume} {2201.07403}} (\bibinfo {year}
  {2022})}\BibitemShut {NoStop}%
\bibitem [{\citenamefont {Tomasch}(1965)}]{tom65}%
  \BibitemOpen
  \bibfield  {author} {\bibinfo {author} {\bibfnamefont {W.~J.}\ \bibnamefont
  {Tomasch}},\ }\bibfield  {{Geometrical Resonance in the Tunneling
  Characteristics of Superconducting Pb}} {\emph {\bibinfo {title}
  {{Geometrical Resonance in the Tunneling Characteristics of Superconducting
  Pb}},\ }}\href {\doibase 10.1103/PhysRevLett.15.672} {\bibfield  {journal}
  {\bibinfo  {journal} {Phys. Rev. Lett.}\ }\textbf {\bibinfo {volume} {15}},\
  \bibinfo {pages} {672} (\bibinfo {year} {1965})}\BibitemShut {NoStop}%
\bibitem [{\citenamefont {Buchholtz}\ \emph {et~al.}(1995)\citenamefont
  {Buchholtz}, \citenamefont {Palumbo}, \citenamefont {Rainer},\ and\
  \citenamefont {Sauls}}]{buc95b}%
  \BibitemOpen
  \bibfield  {author} {\bibinfo {author} {\bibfnamefont {L.}~\bibnamefont
  {Buchholtz}}, \bibinfo {author} {\bibfnamefont {M.}~\bibnamefont {Palumbo}},
  \bibinfo {author} {\bibfnamefont {D.}~\bibnamefont {Rainer}}, \ and\ \bibinfo
  {author} {\bibfnamefont {J.~A.}\ \bibnamefont {Sauls}},\ }\bibfield  {{The
  Effect of Surfaces on the Tunneling Density of States of Anisotropically
  Paired Superconductors}} {\emph {\bibinfo {title} {{The Effect of Surfaces on
  the Tunneling Density of States of Anisotropically Paired Superconductors}},\
  }}\href@noop {} {\bibfield  {journal} {\bibinfo  {journal} {J. Low Temp.
  Phys.}\ }\textbf {\bibinfo {volume} {101}},\ \bibinfo {pages} {1099}
  (\bibinfo {year} {1995})}\BibitemShut {NoStop}%
\bibitem [{\citenamefont {Abrikosov}\ and\ \citenamefont
  {Gorkov}(1959{\natexlab{b}})}]{abr59a}%
  \BibitemOpen
  \bibfield  {author} {\bibinfo {author} {\bibfnamefont {A.~A.}\ \bibnamefont
  {Abrikosov}}\ and\ \bibinfo {author} {\bibfnamefont {L.~P.}\ \bibnamefont
  {Gorkov}},\ }\bibfield  {{On the Theory of Superconducting Alloys I:
  Electrodynamics at $T=0$}} {\emph {\bibinfo {title} {{On the Theory of
  Superconducting Alloys I: Electrodynamics at $T=0$}},\ }}\href@noop {}
  {\bibfield  {journal} {\bibinfo  {journal} {Sov. Phys. JETP}\ }\textbf
  {\bibinfo {volume} {8}},\ \bibinfo {pages} {1090} (\bibinfo {year}
  {1959}{\natexlab{b}})}\BibitemShut {NoStop}%
\bibitem [{\citenamefont {Abramowitz}\ and\ \citenamefont
  {Stegun}(1972)}]{abramowitz72}%
  \BibitemOpen
  \bibfield  {author} {\bibinfo {author} {\bibfnamefont {M.}~\bibnamefont
  {Abramowitz}}\ and\ \bibinfo {author} {\bibfnamefont {I.}~\bibnamefont
  {Stegun}},\ }\href@noop {} {\emph {\bibinfo {title} {Handbook of Mathematical
  Functions}}},\ \bibinfo {edition} {tenth printing}\ ed.\ (\bibinfo
  {publisher} {U.S. Government Printing Office},\ \bibinfo {address}
  {Washington D.C.},\ \bibinfo {year} {1972})\BibitemShut {NoStop}%
\bibitem [{\citenamefont {Xu}\ \emph {et~al.}(1995)\citenamefont {Xu},
  \citenamefont {Yip},\ and\ \citenamefont {Sauls}}]{xu94a}%
  \BibitemOpen
  \bibfield  {author} {\bibinfo {author} {\bibfnamefont {D.}~\bibnamefont
  {Xu}}, \bibinfo {author} {\bibfnamefont {S.~K.}\ \bibnamefont {Yip}}, \ and\
  \bibinfo {author} {\bibfnamefont {J.~A.}\ \bibnamefont {Sauls}},\ }\bibfield
  {{Nonlinear Meissner Effect in Unconventional Superconductors}} {\emph
  {\bibinfo {title} {{Nonlinear Meissner Effect in Unconventional
  Superconductors}},\ }}\href {\doibase 10.1103/PhysRevB.51.16233} {\bibfield
  {journal} {\bibinfo  {journal} {Phys. Rev. B}\ }\textbf {\bibinfo {volume}
  {51}},\ \bibinfo {pages} {16233} (\bibinfo {year} {1995})}\BibitemShut
  {NoStop}%
\bibitem [{\citenamefont {Yip}\ and\ \citenamefont {Sauls}(1992)}]{yip92}%
  \BibitemOpen
  \bibfield  {author} {\bibinfo {author} {\bibfnamefont {S.~K.}\ \bibnamefont
  {Yip}}\ and\ \bibinfo {author} {\bibfnamefont {J.~A.}\ \bibnamefont
  {Sauls}},\ }\bibfield  {{Circular Dichroism and Birefringence in
  Unconventional Superconductors}} {\emph {\bibinfo {title} {{Circular
  Dichroism and Birefringence in Unconventional Superconductors}},\ }}\href
  {\doibase 10.1007/BF01151804} {\bibfield  {journal} {\bibinfo  {journal} {J.
  Low Temp. Phys.}\ }\textbf {\bibinfo {volume} {86}},\ \bibinfo {pages} {257}
  (\bibinfo {year} {1992})}\BibitemShut {NoStop}%
\bibitem [{\citenamefont {Grassellino}\ \emph {et~al.}(2013)\citenamefont
  {Grassellino}, \citenamefont {Romanenko}, \citenamefont {Sergatskov},
  \citenamefont {Melnychuk}, \citenamefont {Trenikhina}, \citenamefont
  {Crawford}, \citenamefont {Rowe}, \citenamefont {Wong}, \citenamefont
  {Khabiboulline},\ and\ \citenamefont {Barkov}}]{gra13}%
  \BibitemOpen
  \bibfield  {author} {\bibinfo {author} {\bibfnamefont {A.}~\bibnamefont
  {Grassellino}}, \bibinfo {author} {\bibfnamefont {A.}~\bibnamefont
  {Romanenko}}, \bibinfo {author} {\bibfnamefont {D.}~\bibnamefont
  {Sergatskov}}, \bibinfo {author} {\bibfnamefont {O.}~\bibnamefont
  {Melnychuk}}, \bibinfo {author} {\bibfnamefont {Y.}~\bibnamefont
  {Trenikhina}}, \bibinfo {author} {\bibfnamefont {A.}~\bibnamefont
  {Crawford}}, \bibinfo {author} {\bibfnamefont {A.}~\bibnamefont {Rowe}},
  \bibinfo {author} {\bibfnamefont {M.}~\bibnamefont {Wong}}, \bibinfo {author}
  {\bibfnamefont {T.}~\bibnamefont {Khabiboulline}}, \ and\ \bibinfo {author}
  {\bibfnamefont {F.}~\bibnamefont {Barkov}},\ }\bibfield  {{Nitrogen and argon
  doping of niobium for superconducting radio frequency cavities: a pathway to
  highly efficient accelerating structures}} {\emph {\bibinfo {title}
  {{Nitrogen and argon doping of niobium for superconducting radio frequency
  cavities: a pathway to highly efficient accelerating structures}},\ }}\href
  {\doibase 10.1088/0953-2048/26/10/102001} {\bibfield  {journal} {\bibinfo
  {journal} {Supercond. Sci. Technol.}\ }\textbf {\bibinfo {volume} {26}},\
  \bibinfo {pages} {102001} (\bibinfo {year} {2013})}\BibitemShut {NoStop}%
\bibitem [{\citenamefont {Bogorad}\ \emph {et~al.}(2019)\citenamefont
  {Bogorad}, \citenamefont {Hook}, \citenamefont {Kahn},\ and\ \citenamefont
  {Soreq}}]{bog19}%
  \BibitemOpen
  \bibfield  {author} {\bibinfo {author} {\bibfnamefont {Z.}~\bibnamefont
  {Bogorad}}, \bibinfo {author} {\bibfnamefont {A.}~\bibnamefont {Hook}},
  \bibinfo {author} {\bibfnamefont {Y.}~\bibnamefont {Kahn}}, \ and\ \bibinfo
  {author} {\bibfnamefont {Y.}~\bibnamefont {Soreq}},\ }\bibfield  {{Probing
  Axionlike Particles and the Axiverse with Superconducting Radio-Frequency
  Cavities}} {\emph {\bibinfo {title} {{Probing Axionlike Particles and the
  Axiverse with Superconducting Radio-Frequency Cavities}},\ }}\href {\doibase
  10.1103/PhysRevLett.123.021801} {\bibfield  {journal} {\bibinfo  {journal}
  {Phys. Rev. Lett.}\ }\textbf {\bibinfo {volume} {123}},\ \bibinfo {pages}
  {021801} (\bibinfo {year} {2019})}\BibitemShut {NoStop}%
\bibitem [{\citenamefont {Gao}\ and\ \citenamefont {Harnik}(2021)}]{cao20}%
  \BibitemOpen
  \bibfield  {author} {\bibinfo {author} {\bibfnamefont {C.}~\bibnamefont
  {Gao}}\ and\ \bibinfo {author} {\bibfnamefont {R.}~\bibnamefont {Harnik}},\
  }\bibfield  {{Axion searches with two superconducting radio-frequency
  cavities}} {\emph {\bibinfo {title} {{Axion searches with two superconducting
  radio-frequency cavities}},\ }}\href {\doibase 10.1007/JHEP07(2021)053}
  {\bibfield  {journal} {\bibinfo  {journal} {J. High Energ. Phys.}\ }\textbf
  {\bibinfo {volume} {2021}},\ \bibinfo {pages} {53} (\bibinfo {year}
  {2021})}\BibitemShut {NoStop}%
\bibitem [{\citenamefont {Padamsee}(2009)}]{pad09}%
  \BibitemOpen
  \bibfield  {author} {\bibinfo {author} {\bibfnamefont {H.}~\bibnamefont
  {Padamsee}},\ }\href {\doibase 10.1002/9783527627172} {\emph {\bibinfo
  {title} {{RF Superconductivity}}}}\ (\bibinfo  {publisher} {Wiley Online
  Library},\ \bibinfo {year} {2009})\ pp.\ \bibinfo {pages}
  {1--454}\BibitemShut {NoStop}%
\bibitem [{\citenamefont {Nambu}(1960)}]{nam60}%
  \BibitemOpen
  \bibfield  {author} {\bibinfo {author} {\bibfnamefont {Y.}~\bibnamefont
  {Nambu}},\ }\bibfield  {{Quasi-Particles and Gauge Invariance in the Theory
  of Superconductivity}} {\emph {\bibinfo {title} {{Quasi-Particles and Gauge
  Invariance in the Theory of Superconductivity}},\ }}\href {\doibase
  10.1103/PhysRev.117.648} {\bibfield  {journal} {\bibinfo  {journal} {Phys.
  Rev.}\ }\textbf {\bibinfo {volume} {117}},\ \bibinfo {pages} {648} (\bibinfo
  {year} {1960})}\BibitemShut {NoStop}%
\end{thebibliography}
%
\end{document}